\documentclass{aastex63}

\usepackage{hyperref,epsfig,graphicx,lineno,longtable, amssymb,multirow}

\received{***}
\revised{***}
\accepted{***}

\shorttitle{Current Helicity}
\shortauthors{Sun et al.}
\graphicspath{{./}{figures/}}

\begin{document}

\title{Current Helicity in Response to Coronal Mass Ejections}

\affiliation{School of Earth and Space Sciences, Peking University, Beijing, 100871, People's Republic of China} % #1
\affiliation{State Key Laboratory of Solar Activity and Space Weather, National Astronomical Observatories, Chinese Academy of Sciences, Beijing 100101, People's Republic of China; \href{liting@nao.cas.cn}{liting@nao.cas.cn}} % #2
\affiliation{Leibniz Institute for Astrophysics Potsdam, An der Sternwarte 16, 14482 Potsdam, Germany} % #3
\affiliation{School of Astronomy and Space Science, University of Chinese Academy of Sciences, Beijing 100049, People's Republic of China} % #4

\author[0000-0001-5657-7587]{Zheng Sun}
\affiliation{School of Earth and Space Sciences, Peking University, Beijing, 100871, People's Republic of China}
\affiliation{Leibniz Institute for Astrophysics Potsdam, An der Sternwarte 16, 14482 Potsdam, Germany}

\author[0000-0001-6655-1743]{Ting Li}
\affiliation{State Key Laboratory of Solar Activity and Space Weather, National Astronomical Observatories, Chinese Academy of Sciences, Beijing 100101, People's Republic of China; \href{liting@nao.cas.cn}{liting@nao.cas.cn}}
\affiliation{School of Astronomy and Space Science, University of Chinese Academy of Sciences, Beijing 100049, People's Republic of China}

\author[0000-0001-9189-1846]{Xinkai Bian}
\affiliation{Shenzhen Key Laboratory of Numerical Prediction for Space Storm, School of Aerospace, Harbin Institute of Technology, Shenzhen 518055, People's Republic of China}

\author[0000-0002-9534-1638]{Yijun Hou}
\affiliation{State Key Laboratory of Solar Activity and Space Weather, National Astronomical Observatories, Chinese Academy of Sciences, Beijing 100101, People's Republic of China; \href{liting@nao.cas.cn}{liting@nao.cas.cn}}
\affiliation{School of Astronomy and Space Science, University of Chinese Academy of Sciences, Beijing 100049, People's Republic of China}

\author[0000-0002-3694-4527]{Ioannis Kontogiannis}
\affiliation{Institute for Particle Physics and Astrophysics, ETH Zürich, Otto-Stern-Weg 5, 8093 Zürich, Switzerland}
\affiliation{Istituto ricerche solari Aldo e Cele Daccò (IRSOL), Faculty of Informatics, Universitá della Svizzera italiana, CH-6605 Locarno, Switzerland}

\author[0000-0002-1349-8720]{Ziqi Wu}
\affiliation{School of Earth and Space Sciences, Peking University, Beijing, 100871, People's Republic of China}
\affiliation{Centre for mathematical Plasma Astrophysics, KU Leuven, Celestijnenlaan 200B bus 2400, B-3001 Leuven, Belgium}

\begin{abstract}
  
Coronal mass ejections (CMEs), powerful solar eruptions with massive plasma ejected into the interplanetary space, are caused by the release of the magnetic free enengy stored in coronal electric currents. Photospheric current helicity, defined as the integral of the product of vertical electric current density and vertical magnetic field ($H_c=\int j_zB_z\ dS$), serves as a key parameter in understanding the eruptions. Using a 3D magnetohydrodynamic model, we identify a current helicity reversal pattern associated with the eruption: a pre-eruption decrease and a post-eruption increase. This helicity reversal is attributed to the redistribution of electric currents: before the eruption, currents concentrate toward the polarity inversion line (PIL); after the eruption they move away from the PIL, consistent with the flare ribbon separation, which is caused by the upward progression reconnection site. To validate this pattern, we conducted an observational analysis of 50 $\geq$M5.0 eruptive flares. The results reveal that 58\% of cases exhibited a pre-eruption decrease and 92\% showed the post-eruption increase in current helicity. Detailed analysis of two cases with this reversal suggests that they share the same current redistribution pattern, consistent with the mechanism identified in the simulations. Moreover, the pre-eruption decrease could be observed clearly even in the long-term evolution of the two cases. Current helicity can serve as an indicator of when electric currents are built up for the subsequent eruption, and it has the potential to predict CMEs to some extent.
\end{abstract}


\section{Introduction } \label{sec:intro}
Solar flares and CMEs are solar eruptions with massive plasma ejected into the interplanetary space, which are driven by magnetic free enengy stored in coronal electric currents (e.g. \citealt{1998GeoRL..25.3019B,2020ApJS..247...21S,2024SoPh..299...93S}). However, critical aspects of how magnetic energy accumulates in coronal magnetic fields and the processes that trigger its rapid release during CMEs remain not fully understood \citep{2008AnGeo..26.3089V,2017ScChD..60.1383C,2024ScChD..67.3765J}. While routine measurements of the photospheric magnetic field are accessible via the Zeeman effect, the magnetic fields in the chromosphere and corona are far weaker, making Zeeman-based detection challenging \citep{2004ApJ...613L.177L,2024arXiv241021568S}. Despite of the advancements in direct and indirect methods to measure chromospheric and coronal magnetic fields \citep{2021ApJ...918L..13C,2020Sci...367..278F,2020Sci...369..694Y,2024Sci...386...76Y}, achieving high spatial and temporal resolution remains difficult. Consequently, studying solar eruptions based on the high-resolution photospheric magnetic field data is essential, which can help us understand the photospheric response to the coronal activities \citep{2018ApJ...864..159W,2024A&A...686A.148S}.

Multiple studies have observed photospheric magnetic field changes during solar flares, which can be categorized into magnetic transients and magnetic imprints \citep{2015RAA....15..145W,2017ApJ...839...67S}. Magnetic transients represent temporary, flare-induced reversals in magnetic polarity, often thought to be artifacts induced by changes in spectral profiles \citep{2003ApJ...599..615Q,2004SoPh..220...81A}. In contrast, magnetic imprints are considered to be direct photospheric feedback from coronal eruptions. These changes on the photosphere are typically expected to be minor due to the large density in the photosphere \citep{2016NatPh..12..998A}, however, in some intense eruptions, photospheric magnetic fields have shown rapid changes (e.g., \citealt{2018ApJ...852...25C,2019ApJ...876..133L,2022arXiv221101911W}). Many studies indicate that horizontal fields near the polarity inversion line (PIL) often increase after an eruption \citep{2010ApJ...716L.195W,2012ApJ...745L..17W,2017ApJ...839...67S}. Such variations are commonly explained by the magnetic implosion conjecture \citep{2000ApJ...531L..75H}, which proposed that the release of free magnetic energy leads to a decreased magnetic pressure. This can cause the coronal field to collapse, making the field in flaring regions more horizontal \citep{zuccarello2017vortex}. Additionally, some studies suggest that compression from downward flows produced by magnetic reconnection can contribute to the increase of horizontal field (e.g., \citealt{2019ApJ...876..133L,2023FrASS..1097672B,2023ApJ...944..215Y}). While the horizontal field near the PIL increases, the vertical field often remains unchanged (e.g., \citealt{2010ApJ...716L.195W,2012ApJ...759...50P}). Moreover, flares can induce sunspot rotation or reversal rotation via Lorentz forces \citep{2016SoPh..291.1159W,2016NatCo...713798B,2016NatCo...713104L}, which \citet{2016NatPh..12..998A} described as ``the tail wagging the dog", referring to coronal evolution driving flows in the more massive photosphere. 

Some magnetic-associated photospheric parameters could also be changed before or after the eruptions. Both magnetic shear and vertical current density have been observed to increase following a flare \citep{2008ApJ...676L..81J,2012ApJ...748...77S,2012ApJ...745L..17W}. This can be explained by the scenario that the solar interior may act as a ``current driver", responding to rapid changes in coronal currents \citep{2000ApJ...545.1089L,2022SoPh..297...59K}. \citet{2008ApJ...686.1397P} found that before X-class flares the magnetic helicity undergoes a monotonic accumulation followed by a constant phase. \citet{2023ApJ...942...27L} observed that helicity flux and energy flux tend to decrease during eruptive flares, whereas they remain stable in confined flares. Additionally, the average photospheric force-free parameter alpha $j_z/B_z$ was found to decrease before eruptive flares and suddenly increase afterward. The horizontal gradient proxy introduced by \citet{2014ApJ...789..107K,2015ApJ...802L..21K} exhibits a specific pre-flare evolutionary pattern. Non-neutralized electric currents and the Ising energy exhibit distinct structural enhancements prior to major flares \citep{2017SoPh..292..159K,2018SoPh..293...96K}.
\citet{2024MNRAS.534..444A} analyzed magnetic winding flux in 30 eruptive events, finding significant topological changes near the PIL preceding or concurrent with CME onset. 

According to the classification by \citet{2014SoPh..289.2091L}, eruption models fall into three primary categories: loss of equilibrium (e.g., \citealt{1979SoPh...64..303H,1990JGR....9511919F,2006PhRvL..96y5002K}), tether-cutting \citep{2001ApJ...552..833M,2018ApJ...869...78C}, and breakout (e.g., \citealt{1999ApJ...510..485A, 2017Natur.544..452W,2023ApJ...953..148S,2024ApJ...962L..38D}) models. Among these, \citet{2015RAA....15..145W} discussed that tether-cutting model has been shown to best accommodate various observational features of the short-term variation associated with the flare. This model describes a process in which coronal arcades undergo shearing motions that lead to magnetic reconnection between sheared arcades, forming an erupting flux rope. Since it was proposed by \citet{2001ApJ...552..833M}, this model remained a conjectural ``cartoon" until \citet{2021NatAs...5.1126J} provided simulations supporting the tether-cutting scenario. Their model shows that photospheric shearing motions alone can create an electric current sheet near the PIL. Once magnetic reconnection initiates, the entire arcades erupt impulsively and creates a fast-expanding flux rope.

Photospheric current helicity (referred to as current helicity hereafter) is defined as the integral of the product of the vertical electric current density and the vertical magnetic field ($H_c=\int j_zB_z\ dS$). While some studies have investigated the time evolution of current helicity in ARs, constrained by the time resolution of the instruments (e.g., \citealt{bao1999survey,deng2001evolution}), few have focused on its short-term variations associated with flares using sub-hour time resolution.
In this study, we aim to investigate the short-term variations in current helicity associated with CMEs. We analyze the variations from the magnetohydrodynamic (MHD) model and observations, which are shown in the Section 2 and Section 3, respectively. The underlying mechanisms are discussed in Section 4. Finally, we briefly summarize our results in Section 5.

\section{Models} \label{sec:model}

We examine the evolution of photospheric parameters in the MHD simulation based on the tether-cutting scenario referred to \citet{2021NatAs...5.1126J}. In this process, shearing flows lead to the formation of highly sheared arcades, which then reconnect with each other and erupt.

\subsection{Methods} \label{sec:model}

The simulation analyzed here is identical to that in \citet{2022ApJ...925L...7B}, except that they simulated three consecutive eruptions whereas we focus solely on the first eruption. MHD equations are solved by the DARE--MHD code \citep{2016ApJ...828...62J}, which is based on the conservation element/solution element (CESE) numerical scheme and incorporates adaptive mesh refinement (AMR) for enhanced resolution \citep{2010ApJ...723..300F, 2010SoPh..267..463J}. The simulation is performed in a Cartesian domain with horizontal extents of \( x, y \in [-270, 270] \)~Mm and a vertical range of \( z \in [0, 540] \)~Mm. The lower boundary at \( z = 0 \), representing the photosphere, is treated as line-tied. A block-structured AMR grid is employed, with a coarsest grid spacing of \( \Delta = 2.88 \)~Mm and finest refinement reaching \( \Delta = 360 \)~km, allowing accurate resolution of small-scale structures such as current sheets and reconnection layers. The simulation begins from a potential magnetic field configuration and is continuously driven by steady rotational motions applied to each magnetic polarity at the bottom boundary.
In our simulation, through photospheric shearing motion alone, the core current sheet initiates an eruption at t = 78 $\tau$ (where the time unit is $\tau$ = 105 s). More details of the model settings can be found in \citet{2022ApJ...925L...7B}.

We use the bottom boundary (Z=0) of the model as the photosphere to calculate the photospheric response to the eruption. All the grids at Z=0 are included in the calculation. Using this photospheric data, we can calculate vertical current density $j_z$ and current helicity density $h_c$:
\begin{equation}
j_z=(\nabla\times \mathbf{B})_z
\end{equation}
\begin{equation}
h_c=j_zB_z
\end{equation}
Then we integrate $B_z$, $j_z$, and $h_c$ over the photosphere (Z=0), we can obtain magnetic flux $\Phi_z$, total vertical current $J_z$, and current helicity $H_c$, respectively:
\begin{equation}
\Phi_z=\int B_z\ dS
\end{equation}
\begin{equation}
J_z=\int j_z \ dS
\end{equation}
\begin{equation}
H_c=\int h_c\ dS
\end{equation}
To illustrate the variation before and after the eruption, we set the eruption onset time as
$t=0$, with times prior to the eruption denoted as $t < 0$.

\subsection{Results} \label{sec:model}
Figure 1(a) shows the overview of the MHD model results. 12 time steps (21 min) are selected before and 8 time steps (14 min) after the eruption to analyze the evolution of photospheric parameters, displayed in panel (b). The parameter evolution indicates that magnetic flux remains unchanged during the eruption, aligning with previous studies \citep{2010ApJ...716L.195W,2012ApJ...759...50P}. The unsigned total current increases by approximately 2\%, due to the shearing motions of the magnetic polarities. Notably, the total current rises more sharply after the eruption, likely due to downward compression from magnetic reconnection \citep{2023FrASS..1097672B}, which enhances the horizontal magnetic field and thus current. Most intriguingly, current helicity shows a reversal: a pre-eruption decrease followed by the post-eruption increase. This behavior cannot be explained by the variation of the magnitude of mean $j_z$ or $B_z$, so we make a further investigation into the 2D maps of the parameters.

Figure 2 presents the pre-eruption, eruption onset and post-eruption states of photospheric distribution of $B_z$, $j_z$, and $h_c$. The $B_z$ maps display a stable bipole configuration and hardly change with time, as shown in panels (a1)-(a3). Regions with strong $j_z$ exhibit a double-J configuration (panels (b1-b3)), which is in accordance with previous studies (e.g., \citealt{2014ApJ...788...60J,2016ApJ...817..156W}). This double-J current structure can be considered as the footprint of quasi-separatrix layer (QSL), which is co-spatial to the flare ribbons \citep{2016ApJ...823...62Z,2019A&A...621A..72A}. 
Comparing panels (b1) and (b2), the current structure evolves from an imperfect J shape to a well-defined J shape before the eruption, suggesting that the current shrinks from the volumetric region to the PIL (also see the animation S1). This variation pattern is similar to findings by \citet{2021NatAs...5.1126J} (see their Figure 1), which is considered to be attributed by the sheared flows. Such a concentration of current distribution corresponds to the increase of the free magnetic energy in the system, which powers the subsequent eruption. After the eruption (comparing panels (b2) and (b3)), it is observed that the double-J current structures are moving away from each other, towards the center of the magnetic poles (marked as red arrows). This trend is similar to flare ribbon separation, a result of the elevated reconnection site as successively larger-scale coronal loops participate in the reconnection \citep{2011SSRv..159...19F}.
Panels (c1)-(c3) depict the current helicity density, $h_c=j_zB_z$, dominated by negative helicity as indicated by the positive $B_z$ field with negative $j_z$ and vice versa. Since $B_z$ remains stable, the variation of $h_c$ is similar to that of the $j_z$: before the eruption, helicity contributed by the magnetic pole region is decreasing due to the current concentration, while after the eruption, the helicity contribution from magnetic poles is increasing because of the current ribbon separation caused by the upward progression of the reconnection site.

Figure 3 further shows $j_z$ evolution via difference images, which effectively highlight $j_z$ variations. Panel (b) represents the $j_z$ difference between the eruption onset and the pre-eruption phases, capturing pre-eruption changes. During this period, the current near the magnetic poles decreases (red arrows in panel (b), while current of the double-J structure increases. This indicates that the current is concentrating from the magnetic poles to the PIL, in accordance with \citet{2021NatAs...5.1126J}. Panel (c) illustrates the $j_z$ difference between the post-eruption and eruption onset phases, used to track post-eruption evolution. In this phase, the double-J structures separate with each other, consistent with the flare ribbon separation. This separation suggests that the current is moving from the PIL back toward the magnetic poles (highlighted by red arrows in panel (c)), a motion opposite to that observed before the eruption.
Since the helicity density is the product of $B_z$ and $j_z$, the contribution from the weak-$B_z$ region (i.e., PIL region here) can be neglected. Therefore, the helicity contribution primarily comes from the magnetic pole regions where $B_z$ is strong. In these regions (inside the orange contour), the current decreases before the eruption (panel (b)) due to current concentration and increases after the eruption (panel (c)) due to the elevated reconnection site.
To quantify this current variation, we calculate the mean $j_z$ in the magnetic poles, shown as the orange curve in panel (d). The curve shows a decrease before and an increase after the eruption, matching the helicity trend in Figure 1(b). This agreement indicates that the helicity reversal originates from the $j_z$ reversal in the strong-$B_z$ region.
We also calculate the mean $j_z$ in the weak-$B_z$ region (the PIL region; blue dashed square in panel (a)), represented by the blue curve in panel (d). It exhibits an opposite trend to that of the strong-$B_z$ region, further confirming the current moving pattern: before the eruption, the current concentrates from the strong-$B_z$ region to the weak-$B_z$ region, while after the eruption, it moves back. The opposite current movements before and after the eruption causes the $j_z$ reversal in the strong-$B_z$ region, thus driving the helicity reversal.

\section{Observations} \label{sec:observation}
Having identified a current helicity evolution pattern from the MHD model, next we study whether this pattern can be detected in observations. To this end, we conducted a statistical analysis to examine photospheric variations in current helicity before and after eruptions.

\subsection{Dataset and Methods} \label{sec:observation}
Our dataset is based on data from the Helioseismic and Magnetic Imager (HMI; \citealt{2012SoPh..275..207S}) on the Solar Dynamics Observatory (SDO; \citealt{2012SoPh..275...17L}), which provides full-disk magnetograms with 0.5\arcsec\ pixel resolution. For our analysis, we primarily use Spaceweather HMI Active Region Patch (SHARP) data \citep{2014SoPh..289.3549B}, which enables automated AR identification and tracking. SHARP data also include Lambert cylindrical equal-area (CEA) projections of magnetic fields, facilitating transformation of [$B_x$,$B_y$,$B_z$] into spherical heliographic coordinates [$B_{\phi}$, $B_{\theta}$, $B_r$] with a 720-second cadence \citep{2013arXiv1309.2392S}. This data allows us to compute both current density and helicity density. We also utilize observations from the Atmospheric Imaging Assembly (AIA) on SDO, specifically 1600~\AA\ images with a resolution of 0.6\arcsec\ and a 24-second cadence, to identify flare locations.

Our study involves 50 samples, focusing on the eruptive flares larger than M5.0-class from June 2010 to June 2024. We calculate current helicity evolution of each event for 10 hours, within 6 hours before the eruption and 4 hours after the eruption. The criteria we use to assemble our catalog are that: (i) ARs remain within 60$^\circ$ of the central meridian; (ii) no other $\geq$M5.0 flares occur in the same AR during the analyzed 10-hour time interval; and (iii) events are associated with a CME, which can be identified in the online CME dataset\footnote{\url{https://cdaw.gsfc.nasa.gov/CME_list/}} and jhelioviewer\footnote{\url{https://www.jhelioviewer.org/}} movies. Details of the selected 50 samples can be seen in Table 1.

We utilize SHARP data patches to analyze AR evolutions. To exclude non-flare-related pixels, we create a flare mask for each event. Following \citet{2017ApJ...846L...6L}, we first calculate the squashing factor Q via NLFFF extrapolation, then compare flare arcades and ribbons in AIA 171~\AA\ and 1600~\AA\ images. Finally, we select high-Q contours to encompass the flare’s arcade and ribbon structures. As demonstrated in \citet{2023ApJ...942...27L}, this method could effectively isolate erupting flux regions. Only pixels within the flare mask will be involved in the calculation. Moreover, we only consider the pixels with a total field strength exceeding 300 G, which is about three times of the transverse field noise level for HMI vector magnetograms \citep{2014SoPh..289.3483H}. We use the same methods as in the numerical model for calculating current and current helicity.
Besides, we identify the PILs of the ARs with a horizontal magnetic gradient above 300 G Mm$^{-1}$.

A case is classified as exhibiting a pre-eruption decrease if a continuous reduction in current helicity persists for over 1 hour (5 time steps) with a relative magnitude greater than 10\%. Similarly, a case is considered to exhibit the post-eruption increase if a continuous increase is observed for over 36 min (3 time steps) and exceeds a 10\% relative magnitude. The evolution pattern of the helicity and the corresponding duration can be seen in Table 1. 

\subsection{Statistical Results} \label{sec:observation}
Analysis of the current helicity evolution in 50 cases shows that 58\% (29/50) exhibit a decrease before eruption, and 92\% (46/50) display an increase after eruption. Here we use the unsigned $H_c$ for the reason that one AR could have positive or negative chirality. We categorized cases based on the presence of pre-eruption decrease (Figure 4) and the red curve is the temporal mean value for the two types. The cases with pre-eruption decrease exhibit the similar trend with that in the model (panel (a)) . Among those with pre-eruption decrease, 11 out of 29 show a steady decline over the six hours preceding the eruption, while some cases begin decreasing just one hour prior. The average decrease magnitude and duration are 27.1\% and 4.1 h, respectively. It should be noted the duration is the lower limit, since many samples exhibit decrease time $\geq$ 6h. For the cases without pre-eruption decrease, some of them remain stable or change irregular, with a few showing a modest pre-eruption increase of less than 10\%. As for the post-eruption decrease, the average magnitude and duration can be 36.93\% and 1.8 h, which is also a lower limit.

\subsection{Case Studies} \label{sec:observation}
To investigate the mechanism of the pre-eruption decrease and post-eruption increase in current helicity from observations, we conduct detailed analyses of two representative cases with this reversal pattern and one case without the pre-eruption decrease.

\subsubsection{Case 1: With Reversal} \label{sec:observation}

Case 1 is an eruptive M9.3 flare on 2011 August 4 in AR 11261, starting at 03:41 UT, accompanied by a halo CME with a linear speed of 1315 km s$^{-1}$. Figure 5 shows the 1600~\AA\ emission of the flare, with bright regions representing flare ribbons. We selected a zoom-in region encompassing the magnetic poles and flaring PILs (white square in panel (a)) to conduct detailed analysis. The evolution of the $|J_z|$, $|\Phi_z|$, and $|H_c|$ are shown in panel (b). Unsigned current $J_z$ shows a slight increase in current before the flare, followed by a sharp increase after the eruption. $\Phi_z$ remains stable, which is consistent with the model result. Notably, $H_c$ decreases by approximately 24.9\% before the eruption and increases 40.1\% after the eruption, similar to the behavior observed in the model (Figure 1(b)). Since $B_z$ is quite stable, the changes in helicity should be associated with the variation of the current, which is similar to that in the model. Therefore, we conduct the similar methods used in the model to analyze the observational cases.

We created difference images and calculated the time-dependent mean $j_z$ shown in Figure 6. Panel (a) shows the pre-eruption $j_z$ distribution, with orange contours indicting the strong-$B_z$ regions. The PIL is marked as the cyan scatters.
Panel (b) depicts the pre-eruption current variations, showing a decrease in the strong-$B_z$ regions (marked as the red arrows) and an increase near the PIL. This suggests that the current is moving from the strong-$B_z$ regions to the weak-$B_z$ regions, consistent with the behavior observed in the model (Figure 3(b)). Panel (c) illustrates the post-eruption $j_z$ variations, revealing a double-ribbon current structure that moves apart from each other (see animation S2). This behavior aligns with the model (Figure 3(c)) and indicates that the current is moving toward the strong-$B_z$ regions. Panel (d) shows the calculated mean $j_z$ evolution in both the strong-$B_z$ and PIL (weak-$B_z$) regions, further quantifying these trends. In strong-$B_z$ areas, $j_z$ decreases before the eruption and increases afterward, while the region near the PIL shows a $j_z$ increases before the eruption and decreases afterward, which are the same as that in the simulation result (Figure 3(d)). Since regions with strong $B_z$ contribute significantly to helicity, the $j_z$ variations in the strong-$B_z$ regions (orange curves in panel (d)) play a major role in determining helicity. Therefore, the mechanism responsible for the helicity reversal pattern aligns with the model findings: the opposite current movement before and after the eruption leads to the reversal of $j_z$ in strong-$B_z$ regions, ultimately causing the helicity reversal.

It should be noted that the decrease (increase) in helicity is essentially attributed to the spatial separation (coincidence) of strong $j_z$ and strong $B_z$, resulting in an decreasing (increasing) $j_zB_z$ product. The magnetic pole exhibits a radially decreasing field strength—being strongest at the center and gradually weakening toward the periphery. Therefore, as the current ribbon moves toward (away from) the center of the magnetic pole, $j_zB_z$ correspondingly increases (decreases).
The term “strong-Bz region” is used as a conceptual simplification to illustrate this behavior. The specific threshold used to define “strong Bz” does not affect the physical interpretation.

\subsubsection{Case 2: With Reversal} \label{sec:observation}

Case 2 involves the eruptive M8.2 flare occuring on 2023 September 20 starting at 14:11 UT, which is located in AR 13435. This event is accompanied by a partial halo CME at the speed of 536 km s$^{-1}$. Figure 7 shows the 1600~\AA\ image of the flaring region, and the mean parameter evolutions are shown in Panel (b). The evolutions of $B_z$, $j_z$ and $h_c$ are similar to that of Case 1, except that the unsigned current shows a decrease before the eruption rather than increase in Case 2.

Figure 8 shows the same analysis as Case 1 in Figure 6. During the pre-eruption phase (panel (b)), a decrease in current density in the strong-$B_z$ region (marked as the red arrows) and an increase in the PIL region are observed, resembling Figure 3(b) and 6(b). In Case 2, a distinct double-ribbon current structure is not identified; however, it is common for the absent of such features in previous observations \citep{2020ApJ...900...38H}. Though we cannot observe a post-eruption, the current density increase in the strong-$B_z$ region still can be observed (marked as the red arrows), which is similar to Figure 6(c). Panel (d) shows the calculated $j_z$ variations in the strong-$B_z$ and weak-$B_z$ regions, which demonstrates the current redistribution pattern again. Therefore, the underlying mechanism of the helicity reversal in the model and the two cases should be the same.

\subsubsection{Case 3: Without Pre-eruption Decrease} \label{sec:observation}

Case 3 corresponds to an eruptive M5.3 flare originating from NOAA AR 11283 on 2011 September 6 at 01:35 UT. The event is associated with a halo CME exhibiting a projected speed of 782 km s$^{-1}$. Figure 9 displays the AIA 1600~\AA\ image of the flaring region, with the temporal evolution of mean parameters shown in panel (b). Unlike the previous two cases and the model, no evident pre-eruption decrease in current helicity is observed. However, a noticeable post-eruption increase in current helicity is still present.

Figure 10 presents the same analysis of current variation as shown in Figure 6 for Case 1. Panels (b) and (d) reveal no clear signature of current concentration prior to the eruption. The green-purple pattern observed in panel (b) within the contour region arises primarily from the displacement of magnetic polarities, rather than a decrease in current. After the eruption, a significant enhancement of $j_z$ is observed in panel (c), accompanied by a $j_z$ decrease in the PIL region (panel (d))). This case is representative of events that do not exhibit pre-eruption current helicity decrease but do show post-eruption enhancements.

\subsection{Long-term Evolution} \label{sec:observation}

Since the pre-eruption decrease in current helicity can be observed for over half of the cases, we intend to investigate whether this trend could be detected much earlier before the eruption. For extended time steps, where magnetic configurations might change significantly, we calculated photospheric evolution for the entire AR rather than using a flare-specific mask. We analyze the current helicity evolution earlier 15 hours before the eruption for the two cases. In Case 1 (shown in Figure 11), a gradual decrease in current helicity appears 15 hours before the first flare (M6.0-class), with an abrupt drop two hours before. Similarly, a $\sim$30\% decrease in helicity is observed six hours before the second flare (M9.3-class). In Case 2 (Figure 12), a $\sim$40\% reduction occurs over 10 hours preceding the first flare (M8.2-class), while the second flare (M8.7-class) shows a smaller decrease. From these two cases, it can be inferred that current helicity may possess some predictive potential. However, it is important to note that the total current helicity integrated across the entire AR may not be effective for complex ARs (e.g., AR 13664). In such cases, the flaring region may represent only a small portion of the entire AR, so the integrated helicity cannot present the flare region clearly.

\section{Discussion} \label{sec:discussion}
In this study, we employed an MHD model to investigate the photospheric changes associated with a solar eruption, focusing on the evolution of current helicity. Our results reveal a helicity reversal through the eruption, namely a pre-eruption decrease and a post-eruption increase. We attribute this reversal variation to the re-distribution of $j_z$ before and after the eruption. The pre-eruption decrease is due to the concentration of currents towards the PIL, while the post-eruption increase is caused by the current separation due to the elevated reconnection site. Consequently, mean $j_z$ in strong-$B_z$ regions decreases before the eruption and increases after the eruption. Since the total current helicity is dominated by contribution from the strong-Bz regions, this $j_z$ reversal in strong-$B_z$ could cause a helicity reversal.
To validate this pattern, we analyzed 50 eruptive flares ($\geq$ M5.0-class) and found that 58\% displayed a pre-eruption helicity decrease, and 92\% showed the post-eruption increase. Detailed analysis of two representative cases with the helicity reversal showed a similar current variation pattern consistent with the model.

The pre-eruption decrease in current helicity is explained as the current concentration towards the PIL, consistent with the findings of \citet{2021NatAs...5.1126J}. In the MHD model, this process corresponds to arcades becoming more sheared due to sheared flows. Consequently, the current structure transitions from a volumetric distribution to a narrow, S-shaped layer (see their Figure 1). It should be noted that the flux rope in the model does not form before the eruption, but during the eruption. However, previous observations indicate that in some cases, flux ropes can be gradually formed prior to eruption \citep{2014Natur.514..465A,2023ApJ...954L..47C}. In those cases, the current concentration might indicate the buildup process of the flux rope. Therefore, the pre-eruption decrease in observations may correspond to either the increased shearing of arcades or the buildup of a flux rope, both of which could contribute to the current concentration required for the subsequent eruption. We also found that 42\% of the cases do not exhibit a pre-eruption decrease in current helicity. This is likely because the sheared arcades or flux rope with enough currents are formed much earlier before the eruption ($\geq$6.0 hours). Upon their formation, they can maintain a stable state until an initiation (e.g., \citealt{2013ApJ...764..125P,2015ApJ...809...34C,2020A&A...642A.109N}), which do not need a gradual current concentration just before the eruption. In this way, current helicity evolution can indicate the time of current accumulation for the eruption, providing insight into when the energy is sufficiently stored. 

The post-eruption increase in current helicity is attributed to the separation of current ribbons towards strong-$B_z$ regions. However, the simulations also show a rapid increase in average current density after eruption, due to the horizontal field increase by compression from downward flows generated during magnetic reconnection, as suggested by \citet{2023FrASS..1097672B}. While this current increase contributes to the helicity increase, the magnitude of the current increase is minor compared to the helicity increase. So we could suggest that the helicity increase is mainly contributed from the separation of ribbons and secondarily from the increase of the horizontal field. Additionally, some of the cases exhibit $\geq$4h increase, which is even larger than the standard duration of solar flares. This may be caused by the background helicity increase of the AR, which might be caused by the AR emergence or continuous shearing motions. 

\citet{2023ApJ...942...27L} conducted a similar study on the evolution of photospheric parameters associated with 15 eruptive flares. They observed that the $B_z$-weighted force-free parameter $\alpha_{weighted}=\frac{\int j_zB_z}{\int B_z^2}$ followed a similar pattern to current helicity in our study, showing a pre-eruption decrease and the post-eruption increase. Looking into the equation of the $B_z$-weighted force-free parameter, it essentially represents current helicity normalized by magnetic flux. Given that magnetic flux remains fairly stable in our simulations and observations, the same trend between $\alpha_{weighted}$ and $H_c$ is expected. While their study provided statistical analysis and speculative explanations, our simulation and detailed case studies offer a more robust interpretation of their results. Additionally, \citet{2023ApJ...942...27L} noted that confined flares lacked a clear pre-eruption decrease and post-eruption increase, which can be explained by the absence of significant current concentration and flare ribbon separation for confined flares \citep{2019ApJ...881..151L}. Previous studies examining the evolution of $\sum j_zB_z$-associated parameters might attribute variations solely to changes in mean magnitudes of current (e.g., \citealt{2012ApJ...752L...9J}). However, our study emphasizes that the redistribution of $j_z$ also plays a critical role in modulating $\sum j_zB_z$. Moreover, some studies used a single time-step value of $\alpha_{weighted}$ or $H_c$ to indicate the non-potentiality of an AR (e.g., \citealt{2015ApJ...804L..28S,2024ApJ...964..159L}). Caution should be taken with this approach, as short-term variations associated with eruptions are significant and cannot be ignored. For instance, in our observations, the helicity decrease can reach up to 42.9\%, while the increases can be as high as 136.5\%. It is more reliable to use the average value over a specific time period for such analyses. Furthermore, for a given AR, a lower value of $\alpha_{weighted}$ or $H_c$ does not necessarily indicate a reduced eruptive potential. This is because current concentration toward PILs can lead to a decrease in helicity while enhancing the likelihood of eruptions.

Flux emergence could cause an increase in $B_z$, thereby influencing the magnitude of $H_c$. To account for this effect, we also calculated the evolution of $H_c/\Phi_z^2$ for each case to eliminate variations caused by flux emergence. Our analysis shows that the trend of $H_c/\Phi_z^2$ is very similar to that of $H_c$ in almost every case (not shown in the paper), indicating that the influence of flux emergence can be neglected. Moreover, we also examined the impact of the flare mask's shape and size on the results. Our tests showed that as long as the modifications were not too substantial, the results for $H_c$ remained nearly unchanged, indicating that $H_c$ is a robust parameter.

``Helicity" is a volume integral requiring integration over a full 3D space, encompassing all directions [$x,\ y,\ z$] to represent the total current helicity. However, photospheric current helicity is a 2D surface integral limited to the photosphere and considers only the z-direction component, which captures only a portion of total current helicity. Therefore, we have to be cautious that the short-term variations of photospheric current helicity can not indicate the total current helicity variation in the three-dimension domain, for example, that in corona. In our simulations, we also calculated the total current helicity in the 3D domain and found that the trend is totally different from that of photosphere (which is not shown in the paper). For observations, investigating helicity variations in the 3D domain requires the evolutionary 3D extrapolated coronal magnetic field data, similar to works by \citet{2023ApJ...942...27L} and \citet{2025arXiv250105116T}, which is part of our future work. Moreover, it should be noted that this pre-eruption current concentration pattern is not exclusive to the tether-cutting scenario. Any process involving the gradual formation of a coherent flux rope, e.g., those initiated from a pre-existing seed flux rope \citep{2021ApJ...909...91K,2022ApJ...934..103H}, may exhibit a similar evolution in current helicity.

\section{Conclusion} \label{sec:discussion}
In this work, we conduct an MHD model analysis and an observational statistics to explore the photospheric current helicity ($H_c=\int j_zB_z\ dS$) evolution in response to CMEs. Our MHD results show a reversal of current helicity, with a decrease before and an increase after the eruption. The variation of the helicity is not due to the changes in magnitudes of mean $j_z$ or $B_z$. However, it is caused by the redistribution of the current, which causes the variations of $j_z$ in strong-$B_z$ regions. The pre-eruption decrease is attributed to the concentration of currents towards the PIL, while the post-eruption increase mainly results from the separation of the current ribbons caused by the upward progression of the reconnection site. 

Observations involving 50 eruptive flares reveal that 58\% of the cases exhibit a pre-eruption decrease, and 92\% show the post-eruption increase. The current concentration and ribbon separation can also be observed in the selected two cases with the reversal pattern, which indicates that the mechanism of this reversal pattern is similar to that in the simulation. The pre-eruption current helicity evolution could offer clues for when the current and energy for the eruption are prepared.
Moreoever, this variation pattern is evident even in the long-term evolution of ARs for some cases, indicating that current helicity may have predictive potential for CMEs to some extent.

\acknowledgments
We thank Prof. Chaowei Jiang, Prof. Brian Welsch, Dr. Julián David Alvarado-Gómez and Ling Xin for fruitful discussions. HMI and AIA are payloads onboard \emph{SDO}, a mission of NASA's Living With a Star Program. This work is supported by the B-type Strategic Priority Program of the Chinese Academy of Sciences (grant No. XDB0560000), the National Natural Science Foundations of China (12222306, 12273060, 12303057), the National Key R\&D Program of China No. 2022YFF0503800, and the Youth Innovation Promotion Association of CAS (2023063), and the Specialized Research Fund for State Key Laboratory of Solar Activity and Space Weather. IK was supported by the Deutsche Forschungsgemeinschaft (DFG) project number KO 6283/2-1.

\begin{table}
\centering
\caption{List of 50 Eruptive Flares in Our Dataset}

\begin{tabular}{cccccccc}

\hline
\hline
 \multirow{2}{*}{AR Number} & \multirow{2}{*}{Flare Class} & \multirow{2}{*}{Flare Start Time} &  \multicolumn{2}{c}{Pre-eruption Decrease}  & \multicolumn{2}{c}{Post-eruption Increase}  \\ & & & Relative Magnitude (\%) & Duration (h)& Relative Magnitude (\%) & Duration (h)\\
 \hline

11158     & M6.6       & 2011-02-13 17:28:00               & -                  & -                              & 30.0                 & 1.0                 \\  
11158     & X2.2       & 2011-02-15 01:44:00              & -              & -                           & 10.8               & 0.6               \\  
11261     & M6.0       & 2011-08-03 13:17:00              & 24.9               & $\geq$6.0                             & 40.1               & 1.6               \\  
11261     & M9.3       & 2011-08-04 03:41:00              & 11.8               & 1.2                            & 41.1               & $\geq$4.0                \\  
11283     & M5.3       & 2011-09-06 01:35:00               & -                  & -                              & 68.6               & 2.0                 \\  
11283     & X2.1       & 2011-09-06 22:12:00               & -                  & -                              & 30.2               & $\geq$4.0                \\  
11283     & X1.8       & 2011-09-07 22:32:00               & -                  & -                              & 31.1               & 1.2               \\  
11402     & M8.7       & 2012-01-23 03:38:00              & 10.4               & 2.0                              & 31.1               & 3.0                 \\  
11429     & X5.4       & 2012-03-07 00:02:00               & -                  & -                              & 36.5               & 1.4               \\  
11429     & M6.3       & 2012-03-09 03:22:00               & -                  & -                               & -                  & -                 \\  
11429     & M8.4       & 2012-03-10 17:15:00              & 18.1               & $\geq$6.0                             & 57.8               & 1.4               \\  
11515     & M5.6       & 2012-07-02 10:43:00              & 76.1               & $\geq$6.0                             & 136.5              & 0.8               \\  
11515     & M5.3       & 2012-07-04 09:47:00              & 53.6               & 1                              & 86.4               & 1.8               \\  
11520     & X1.4       & 2012-07-12 15:37:00               & -                  & -                              & 12.5               & 1.0                 \\  
11613     & M6.0       & 2012-11-13 01:58:00              & 33.0               & 1.2                            & 41.3               & 0.8               \\  
11719     & M6.5       & 2013-04-11 06:55:00               & -                  & -                              & 25.0                 & $\geq$4.0                \\  
11877 & M9.3 & 2013-10-24 00:21:00   & 32.2 & 4.2   & 27.8 & 1.6 \\  
11890 & X3.3 & 2013-11-05 22:07:00   & 42.2 & $\geq$6.0   & 19.7 & 0.6 \\ 
11890 & X1.1 & 2013-11-08 04:20:00   & 16.7 & 3.0   & 20.5 & 2.2 \\  
11890 & X1.1 & 2013-11-10 05:08:00   & 32.9 & $\geq$6.0   & 42.0 & 2.0 \\ 
11936 & M6.4 & 2013-12-31 21:45:00   & 29.9 & $\geq$6.0   & 23.2 & 0.4 \\  
12017 & X1.0 & 2014-03-29 17:35:00   & 24.0 & 4.0   & 23.7 & 2.0 \\  
12036 & M7.3 & 2014-04-18 12:31:00   & 38.6 & $\geq$6.0   & 111.7 & $\geq$4.0 \\  
12158 & X1.6 & 2014-09-10 17:21:00   & 19.7 & $\geq$6.0   & 31.5 & 1.6 \\  
12173 & M5.1 & 2014-09-28 02:39:00   & - & -   & - & - \\  
12205 & X1.6 & 2014-11-07 16:53:00   & 15.9 & 4.0   & 34.3 & 3.0 \\  
12241 & M6.9 & 2014-12-18 21:41:00   & 12.0 & 4.0   & 11.0 & 1.2 \\  
12242 & M8.7 & 2014-12-17 04:25:00   & 10.1 & 3.0   & 10.5 & 0.8 \\  
12242 & X1.8 & 2014-12-20 00:11:00   & - & -   & 51.7 & $\geq$4.0 \\  
12297 & X2.1 & 2015-03-11 16:11:00   & - & -   & 27.7 & $\geq$4.0 \\  
12297 & M5.1 & 2015-03-10 03:19:00   & 29.3 & 1.8   & 16.7 & 1.2 \\ 
12371 & M6.5 & 2015-06-22 17:39:00   & - & -   & 54.8 & 2.4 \\  
12371 & M7.9 & 2015-06-25 08:02:00   & 18.3 & $\geq$6.0   & 36.0 & 1.2 \\ 
12673 & M5.5 & 2017-09-04 20:28:00   & 13.1 & 1.8   & 22.8 & 0.8 \\  
12673 & X9.3 & 2017-09-06 11:53:00   & 42.9 & 1.6   & 44.7 & 0.8 \\  
12887 & X1.0 & 2021-10-28 15:17:00   & - & -   & 46.1 & 1.8 \\  
12975 & X1.3 & 2022-03-30 17:21:00   & - & -   & 14.3 & 1.6 \\  
12975 & M9.6 & 2022-03-31 18:17:00   & 19.0 & 5.0   & 14.0 & 1.0 \\  
13004 & M5.7 & 2022-05-04 08:45:00   & - & -   & 22.3 & 1.0 \\  
13006 & X1.5 & 2022-05-10 13:50:00   & - & -   & 20.5 & 1.2 \\  
13229 & M6.4 & 2023-02-25 18:40:00   & 37.2 & 5.0   & - & - \\  
13296 & M6.5 & 2023-05-09 03:42:00   & - & -   & 23.4 & 0.6 \\  
13296 & M5.0 & 2023-05-09 20:32:00   & - & -   & 92.0 & 2.0 \\  
13435 & M8.2 & 2023-09-20 14:11:00   & 19.7 & 4.0   & 21.3 & 2.8 \\  
13435 & M8.7 & 2023-09-21 12:42:00   & 17.6 & 2.0   & - & - \\  
13500 & M9.8 & 2023-11-28 19:35:00   & 41.3 & $\geq$6.0   & 28.0 & 0.8 \\  
13664 & X2.3 & 2024-05-09 08:45:00   & - & -   & 23.6 & 0.6 \\  
13664 & X4.0 & 2024-05-10 06:27:00   & 32.8 & 2   & 43.1 & $\geq$4.0 \\  
13664 & M6.0 & 2024-05-10 13:58:00   & - & -   & 15.9 & $\geq$4.0 \\  
13697 & X1.0 & 2024-06-01 18:24:00   & 13.5 & $\geq$6.0   & 15.0 & 0.8 \\  
\hline
\hline
\end{tabular}

*Note: The ``$\geq$6" (``$\geq$4") indicates that the decrease (increase) persists throughout the entire observation period.
\label{tab:flare-summary-symbols}
\end{table}
\newpage

\bibliographystyle{aasjournal}
\bibliography{ref}

\begin{thebibliography}{}
\expandafter\ifx\csname natexlab\endcsname\relax\def\natexlab#1{#1}\fi
\providecommand{\url}[1]{\href{#1}{#1}}
\providecommand{\dodoi}[1]{doi:~\href{http://doi.org/#1}{\nolinkurl{#1}}}
\providecommand{\doeprint}[1]{\href{http://ascl.net/#1}{\nolinkurl{http://ascl.net/#1}}}
\providecommand{\doarXiv}[1]{\href{https://arxiv.org/abs/#1}{\nolinkurl{https://arxiv.org/abs/#1}}}

\bibitem[{{Abramenko} \& {Baranovsky}(2004)}]{2004SoPh..220...81A}
{Abramenko}, V.~I., \& {Baranovsky}, E.~A. 2004, \solphys, 220, 81, \dodoi{10.1023/B:sola.0000023432.42145.b0}

\bibitem[{{Amari} {et~al.}(2014){Amari}, {Canou}, \& {Aly}}]{2014Natur.514..465A}
{Amari}, T., {Canou}, A., \& {Aly}, J.-J. 2014, \nat, 514, 465, \dodoi{10.1038/nature13815}

\bibitem[{{Antiochos} {et~al.}(1999){Antiochos}, {DeVore}, \& {Klimchuk}}]{1999ApJ...510..485A}
{Antiochos}, S.~K., {DeVore}, C.~R., \& {Klimchuk}, J.~A. 1999, \apj, 510, 485, \dodoi{10.1086/306563}

\bibitem[{{Aslam} {et~al.}(2024){Aslam}, {MacTaggart}, {Williams}, {Fletcher}, \& {Romano}}]{2024MNRAS.534..444A}
{Aslam}, O.~P.~M., {MacTaggart}, D., {Williams}, T., {Fletcher}, L., \& {Romano}, P. 2024, \mnras, 534, 444, \dodoi{10.1093/mnras/stae2110}

\bibitem[{{Aulanier}(2016)}]{2016NatPh..12..998A}
{Aulanier}, G. 2016, Nature Physics, 12, 998, \dodoi{10.1038/nphys3938}

\bibitem[{{Aulanier} \& {Dud{\'\i}k}(2019)}]{2019A&A...621A..72A}
{Aulanier}, G., \& {Dud{\'\i}k}, J. 2019, \aap, 621, A72, \dodoi{10.1051/0004-6361/201834221}

\bibitem[{Bao {et~al.}(1999)Bao, Zhang, Ai, \& Zhang}]{bao1999survey}
Bao, S., Zhang, H., Ai, G., \& Zhang, M. 1999, Astronomy and Astrophysics Supplement Series, 139, 311

\bibitem[{{Bi} {et~al.}(2016){Bi}, {Jiang}, {Yang}, {Hong}, {Li}, {Yang}, \& {Xu}}]{2016NatCo...713798B}
{Bi}, Y., {Jiang}, Y., {Yang}, J., {et~al.} 2016, Nature Communications, 7, 13798, \dodoi{10.1038/ncomms13798}

\bibitem[{{Bian} \& {Jiang}(2023)}]{2023FrASS..1097672B}
{Bian}, X., \& {Jiang}, C. 2023, Frontiers in Astronomy and Space Sciences, 10, 1097672, \dodoi{10.3389/fspas.2023.1097672}

\bibitem[{{Bian} {et~al.}(2022){Bian}, {Jiang}, {Feng}, {Zuo}, \& {Wang}}]{2022ApJ...925L...7B}
{Bian}, X., {Jiang}, C., {Feng}, X., {Zuo}, P., \& {Wang}, Y. 2022, \apjl, 925, L7, \dodoi{10.3847/2041-8213/ac4980}

\bibitem[{{Bobra} {et~al.}(2014){Bobra}, {Sun}, {Hoeksema}, {Turmon}, {Liu}, {Hayashi}, {Barnes}, \& {Leka}}]{2014SoPh..289.3549B}
{Bobra}, M.~G., {Sun}, X., {Hoeksema}, J.~T., {et~al.} 2014, \solphys, 289, 3549, \dodoi{10.1007/s11207-014-0529-3}

\bibitem[{{Brueckner} {et~al.}(1998){Brueckner}, {Delaboudiniere}, {Howard}, {Paswaters}, {St. Cyr}, {Schwenn}, {Lamy}, {Simnett}, {Thompson}, \& {Wang}}]{1998GeoRL..25.3019B}
{Brueckner}, G.~E., {Delaboudiniere}, J.~P., {Howard}, R.~A., {et~al.} 1998, \grl, 25, 3019, \dodoi{10.1029/98GL00704}

\bibitem[{{Castellanos Dur{\'a}n} {et~al.}(2018){Castellanos Dur{\'a}n}, {Kleint}, \& {Calvo-Mozo}}]{2018ApJ...852...25C}
{Castellanos Dur{\'a}n}, J.~S., {Kleint}, L., \& {Calvo-Mozo}, B. 2018, \apj, 852, 25, \dodoi{10.3847/1538-4357/aa9d37}

\bibitem[{{Chen} {et~al.}(2018){Chen}, {Duan}, {Yang}, {Yang}, \& {Dai}}]{2018ApJ...869...78C}
{Chen}, H., {Duan}, Y., {Yang}, J., {Yang}, B., \& {Dai}, J. 2018, \apj, 869, 78, \dodoi{10.3847/1538-4357/aaead1}

\bibitem[{{Chen} {et~al.}(2021){Chen}, {Liu}, {Tian}, {Bai}, {Jin}, {Li}, {Yang}, {Yang}, \& {Deng}}]{2021ApJ...918L..13C}
{Chen}, Y., {Liu}, X., {Tian}, H., {et~al.} 2021, \apjl, 918, L13, \dodoi{10.3847/2041-8213/ac1e9a}

\bibitem[{{Cheng} {et~al.}(2017){Cheng}, {Guo}, \& {Ding}}]{2017ScChD..60.1383C}
{Cheng}, X., {Guo}, Y., \& {Ding}, M. 2017, Science China Earth Sciences, 60, 1383, \dodoi{10.1007/s11430-017-9074-6}

\bibitem[{{Cheng} {et~al.}(2023){Cheng}, {Xing}, {Aulanier}, {Solanki}, {Peter}, \& {Ding}}]{2023ApJ...954L..47C}
{Cheng}, X., {Xing}, C., {Aulanier}, G., {et~al.} 2023, \apjl, 954, L47, \dodoi{10.3847/2041-8213/acf3e4}

\bibitem[{{Chintzoglou} {et~al.}(2015){Chintzoglou}, {Patsourakos}, \& {Vourlidas}}]{2015ApJ...809...34C}
{Chintzoglou}, G., {Patsourakos}, S., \& {Vourlidas}, A. 2015, \apj, 809, 34, \dodoi{10.1088/0004-637X/809/1/34}

\bibitem[{Deng {et~al.}(2001)Deng, Wang, Yan, \& Zhang}]{deng2001evolution}
Deng, Y., Wang, J., Yan, Y., \& Zhang, J. 2001, Solar Physics, 204, 11

\bibitem[{{Duan} {et~al.}(2024){Duan}, {Tian}, {Chen}, {Shen}, {Sun}, {Hou}, \& {Li}}]{2024ApJ...962L..38D}
{Duan}, Y., {Tian}, H., {Chen}, H., {et~al.} 2024, \apjl, 962, L38, \dodoi{10.3847/2041-8213/ad24f3}

\bibitem[{{Feng} {et~al.}(2010){Feng}, {Yang}, {Xiang}, {Wu}, {Zhou}, \& {Zhong}}]{2010ApJ...723..300F}
{Feng}, X., {Yang}, L., {Xiang}, C., {et~al.} 2010, \apj, 723, 300, \dodoi{10.1088/0004-637X/723/1/300}

\bibitem[{{Fleishman} {et~al.}(2020){Fleishman}, {Gary}, {Chen}, {Kuroda}, {Yu}, \& {Nita}}]{2020Sci...367..278F}
{Fleishman}, G.~D., {Gary}, D.~E., {Chen}, B., {et~al.} 2020, Science, 367, 278, \dodoi{10.1126/science.aax6874}

\bibitem[{{Fletcher} {et~al.}(2011){Fletcher}, {Dennis}, {Hudson}, {Krucker}, {Phillips}, {Veronig}, {Battaglia}, {Bone}, {Caspi}, {Chen}, {Gallagher}, {Grigis}, {Ji}, {Liu}, {Milligan}, \& {Temmer}}]{2011SSRv..159...19F}
{Fletcher}, L., {Dennis}, B.~R., {Hudson}, H.~S., {et~al.} 2011, \ssr, 159, 19, \dodoi{10.1007/s11214-010-9701-8}

\bibitem[{{Forbes}(1990)}]{1990JGR....9511919F}
{Forbes}, T.~G. 1990, \jgr, 95, 11919, \dodoi{10.1029/JA095iA08p11919}

\bibitem[{{He} {et~al.}(2022){He}, {Hu}, {Jiang}, {Qiu}, \& {Prasad}}]{2022ApJ...934..103H}
{He}, W., {Hu}, Q., {Jiang}, C., {Qiu}, J., \& {Prasad}, A. 2022, \apj, 934, 103, \dodoi{10.3847/1538-4357/ac78df}

\bibitem[{{He} {et~al.}(2020){He}, {Liu}, {Liu}, {Chen}, {Wang}, \& {Wang}}]{2020ApJ...900...38H}
{He}, Y., {Liu}, R., {Liu}, L., {et~al.} 2020, \apj, 900, 38, \dodoi{10.3847/1538-4357/aba52a}

\bibitem[{{Hoeksema} {et~al.}(2014){Hoeksema}, {Liu}, {Hayashi}, {Sun}, {Schou}, {Couvidat}, {Norton}, {Bobra}, {Centeno}, {Leka}, {Barnes}, \& {Turmon}}]{2014SoPh..289.3483H}
{Hoeksema}, J.~T., {Liu}, Y., {Hayashi}, K., {et~al.} 2014, \solphys, 289, 3483, \dodoi{10.1007/s11207-014-0516-8}

\bibitem[{{Hood} \& {Priest}(1979)}]{1979SoPh...64..303H}
{Hood}, A.~W., \& {Priest}, E.~R. 1979, \solphys, 64, 303, \dodoi{10.1007/BF00151441}

\bibitem[{{Hudson}(2000)}]{2000ApJ...531L..75H}
{Hudson}, H.~S. 2000, \apjl, 531, L75, \dodoi{10.1086/312516}

\bibitem[{{Janvier} {et~al.}(2014){Janvier}, {Aulanier}, {Bommier}, {Schmieder}, {D{\'e}moulin}, \& {Pariat}}]{2014ApJ...788...60J}
{Janvier}, M., {Aulanier}, G., {Bommier}, V., {et~al.} 2014, \apj, 788, 60, \dodoi{10.1088/0004-637X/788/1/60}

\bibitem[{{Jiang}(2024)}]{2024ScChD..67.3765J}
{Jiang}, C. 2024, Science China Earth Sciences, 67, 3765, \dodoi{10.1007/s11430-023-1402-3}

\bibitem[{{Jiang} {et~al.}(2010){Jiang}, {Feng}, {Zhang}, \& {Zhong}}]{2010SoPh..267..463J}
{Jiang}, C., {Feng}, X., {Zhang}, J., \& {Zhong}, D. 2010, \solphys, 267, 463, \dodoi{10.1007/s11207-010-9649-6}

\bibitem[{{Jiang} {et~al.}(2016){Jiang}, {Wu}, {Yurchyshyn}, {Wang}, {Feng}, \& {Hu}}]{2016ApJ...828...62J}
{Jiang}, C., {Wu}, S.~T., {Yurchyshyn}, V., {et~al.} 2016, \apj, 828, 62, \dodoi{10.3847/0004-637X/828/1/62}

\bibitem[{{Jiang} {et~al.}(2021){Jiang}, {Feng}, {Liu}, {Yan}, {Hu}, {Moore}, {Duan}, {Cui}, {Zuo}, {Wang}, \& {Wei}}]{2021NatAs...5.1126J}
{Jiang}, C., {Feng}, X., {Liu}, R., {et~al.} 2021, Nature Astronomy, 5, 1126, \dodoi{10.1038/s41550-021-01414-z}

\bibitem[{{Jing} {et~al.}(2012){Jing}, {Park}, {Liu}, {Lee}, {Wiegelmann}, {Xu}, {Deng}, \& {Wang}}]{2012ApJ...752L...9J}
{Jing}, J., {Park}, S.-H., {Liu}, C., {et~al.} 2012, \apjl, 752, L9, \dodoi{10.1088/2041-8205/752/1/L9}

\bibitem[{{Jing} {et~al.}(2008){Jing}, {Wiegelmann}, {Suematsu}, {Kubo}, \& {Wang}}]{2008ApJ...676L..81J}
{Jing}, J., {Wiegelmann}, T., {Suematsu}, Y., {Kubo}, M., \& {Wang}, H. 2008, \apjl, 676, L81, \dodoi{10.1086/587058}

\bibitem[{{Kazachenko} {et~al.}(2022){Kazachenko}, {Albelo-Corchado}, {Tamburri}, \& {Welsch}}]{2022SoPh..297...59K}
{Kazachenko}, M.~D., {Albelo-Corchado}, M.~F., {Tamburri}, C.~A., \& {Welsch}, B.~T. 2022, \solphys, 297, 59, \dodoi{10.1007/s11207-022-01987-6}

\bibitem[{{Kliem} {et~al.}(2021){Kliem}, {Lee}, {Liu}, {White}, {Liu}, \& {Masuda}}]{2021ApJ...909...91K}
{Kliem}, B., {Lee}, J., {Liu}, R., {et~al.} 2021, \apj, 909, 91, \dodoi{10.3847/1538-4357/abda37}

\bibitem[{{Kliem} \& {T{\"o}r{\"o}k}(2006)}]{2006PhRvL..96y5002K}
{Kliem}, B., \& {T{\"o}r{\"o}k}, T. 2006, \prl, 96, 255002, \dodoi{10.1103/PhysRevLett.96.255002}

\bibitem[{{Kontogiannis} {et~al.}(2017){Kontogiannis}, {Georgoulis}, {Park}, \& {Guerra}}]{2017SoPh..292..159K}
{Kontogiannis}, I., {Georgoulis}, M.~K., {Park}, S.-H., \& {Guerra}, J.~A. 2017, \solphys, 292, 159, \dodoi{10.1007/s11207-017-1185-1}

\bibitem[{{Kontogiannis} {et~al.}(2018){Kontogiannis}, {Georgoulis}, {Park}, \& {Guerra}}]{2018SoPh..293...96K}
---. 2018, \solphys, 293, 96, \dodoi{10.1007/s11207-018-1317-2}

\bibitem[{{Kors{\'o}s} {et~al.}(2014){Kors{\'o}s}, {Baranyi}, \& {Ludm{\'a}ny}}]{2014ApJ...789..107K}
{Kors{\'o}s}, M.~B., {Baranyi}, T., \& {Ludm{\'a}ny}, A. 2014, \apj, 789, 107, \dodoi{10.1088/0004-637X/789/2/107}

\bibitem[{{Kors{\'o}s} {et~al.}(2015){Kors{\'o}s}, {Ludm{\'a}ny}, {Erd{\'e}lyi}, \& {Baranyi}}]{2015ApJ...802L..21K}
{Kors{\'o}s}, M.~B., {Ludm{\'a}ny}, A., {Erd{\'e}lyi}, R., \& {Baranyi}, T. 2015, \apjl, 802, L21, \dodoi{10.1088/2041-8205/802/2/L21}

\bibitem[{{Lemen} {et~al.}(2012){Lemen}, {Title}, {Akin}, {Boerner}, {Chou}, {Drake}, {Duncan}, {Edwards}, {Friedlaender}, {Heyman}, {Hurlburt}, {Katz}, {Kushner}, {Levay}, {Lindgren}, {Mathur}, {McFeaters}, {Mitchell}, {Rehse}, {Schrijver}, {Springer}, {Stern}, {Tarbell}, {Wuelser}, {Wolfson}, {Yanari}, {Bookbinder}, {Cheimets}, {Caldwell}, {Deluca}, {Gates}, {Golub}, {Park}, {Podgorski}, {Bush}, {Scherrer}, {Gummin}, {Smith}, {Auker}, {Jerram}, {Pool}, {Soufli}, {Windt}, {Beardsley}, {Clapp}, {Lang}, \& {Waltham}}]{2012SoPh..275...17L}
{Lemen}, J.~R., {Title}, A.~M., {Akin}, D.~J., {et~al.} 2012, \solphys, 275, 17, \dodoi{10.1007/s11207-011-9776-8}

\bibitem[{{Li} {et~al.}(2019){Li}, {Liu}, {Hou}, \& {Zhang}}]{2019ApJ...881..151L}
{Li}, T., {Liu}, L., {Hou}, Y., \& {Zhang}, J. 2019, \apj, 881, 151, \dodoi{10.3847/1538-4357/ab3121}

\bibitem[{{Li} {et~al.}(2024){Li}, {Zheng}, {Li}, {Hou}, {Li}, {Zhang}, \& {Chen}}]{2024ApJ...964..159L}
{Li}, T., {Zheng}, Y., {Li}, X., {et~al.} 2024, \apj, 964, 159, \dodoi{10.3847/1538-4357/ad2e90}

\bibitem[{{Lin} {et~al.}(2004){Lin}, {Kuhn}, \& {Coulter}}]{2004ApJ...613L.177L}
{Lin}, H., {Kuhn}, J.~R., \& {Coulter}, R. 2004, \apjl, 613, L177, \dodoi{10.1086/425217}

\bibitem[{{Liu} {et~al.}(2016){Liu}, {Xu}, {Cao}, {Deng}, {Lee}, {Hudson}, {Gary}, {Wang}, {Jing}, \& {Wang}}]{2016NatCo...713104L}
{Liu}, C., {Xu}, Y., {Cao}, W., {et~al.} 2016, Nature Communications, 7, 13104, \dodoi{10.1038/ncomms13104}

\bibitem[{{Liu} {et~al.}(2017){Liu}, {Sun}, {T{\"o}r{\"o}k}, {Titov}, \& {Leake}}]{2017ApJ...846L...6L}
{Liu}, Y., {Sun}, X., {T{\"o}r{\"o}k}, T., {Titov}, V.~S., \& {Leake}, J.~E. 2017, \apjl, 846, L6, \dodoi{10.3847/2041-8213/aa861e}

\bibitem[{{Liu} {et~al.}(2023){Liu}, {Welsch}, {Valori}, {Georgoulis}, {Guo}, {Pariat}, {Park}, \& {Thalmann}}]{2023ApJ...942...27L}
{Liu}, Y., {Welsch}, B.~T., {Valori}, G., {et~al.} 2023, \apj, 942, 27, \dodoi{10.3847/1538-4357/aca3a6}

\bibitem[{{Longcope} \& {Forbes}(2014)}]{2014SoPh..289.2091L}
{Longcope}, D.~W., \& {Forbes}, T.~G. 2014, \solphys, 289, 2091, \dodoi{10.1007/s11207-013-0464-8}

\bibitem[{{Longcope} \& {Welsch}(2000)}]{2000ApJ...545.1089L}
{Longcope}, D.~W., \& {Welsch}, B.~T. 2000, \apj, 545, 1089, \dodoi{10.1086/317846}

\bibitem[{{Lu} {et~al.}(2019){Lu}, {Cao}, {Jin}, {Zhang}, {Ding}, \& {Guo}}]{2019ApJ...876..133L}
{Lu}, Z., {Cao}, W., {Jin}, G., {et~al.} 2019, \apj, 876, 133, \dodoi{10.3847/1538-4357/ab16d4}

\bibitem[{{Moore} {et~al.}(2001){Moore}, {Sterling}, {Hudson}, \& {Lemen}}]{2001ApJ...552..833M}
{Moore}, R.~L., {Sterling}, A.~C., {Hudson}, H.~S., \& {Lemen}, J.~R. 2001, \apj, 552, 833, \dodoi{10.1086/320559}

\bibitem[{{Nindos} {et~al.}(2020){Nindos}, {Patsourakos}, {Vourlidas}, {Cheng}, \& {Zhang}}]{2020A&A...642A.109N}
{Nindos}, A., {Patsourakos}, S., {Vourlidas}, A., {Cheng}, X., \& {Zhang}, J. 2020, \aap, 642, A109, \dodoi{10.1051/0004-6361/202038832}

\bibitem[{{Park} {et~al.}(2008){Park}, {Lee}, {Choe}, {Chae}, {Jeong}, {Yang}, {Jing}, \& {Wang}}]{2008ApJ...686.1397P}
{Park}, S.-H., {Lee}, J., {Choe}, G.~S., {et~al.} 2008, \apj, 686, 1397, \dodoi{10.1086/591117}

\bibitem[{{Patsourakos} {et~al.}(2013){Patsourakos}, {Vourlidas}, \& {Stenborg}}]{2013ApJ...764..125P}
{Patsourakos}, S., {Vourlidas}, A., \& {Stenborg}, G. 2013, \apj, 764, 125, \dodoi{10.1088/0004-637X/764/2/125}

\bibitem[{{Petrie}(2012)}]{2012ApJ...759...50P}
{Petrie}, G.~J.~D. 2012, \apj, 759, 50, \dodoi{10.1088/0004-637X/759/1/50}

\bibitem[{{Qiu} \& {Gary}(2003)}]{2003ApJ...599..615Q}
{Qiu}, J., \& {Gary}, D.~E. 2003, \apj, 599, 615, \dodoi{10.1086/379146}

\bibitem[{{Schad} {et~al.}(2024){Schad}, {Petrie}, {Kuhn}, {Fehlmann}, {Rimmele}, {Tritschler}, {Woeger}, {Scholl}, {Williams}, {Harrington}, {Paraschiv}, \& {Szente}}]{2024arXiv241021568S}
{Schad}, T.~A., {Petrie}, G. J.~D., {Kuhn}, J.~R., {et~al.} 2024, arXiv e-prints, arXiv:2410.21568, \dodoi{10.48550/arXiv.2410.21568}

\bibitem[{{Scherrer} {et~al.}(2012){Scherrer}, {Schou}, {Bush}, {Kosovichev}, {Bogart}, {Hoeksema}, {Liu}, {Duvall}, {Zhao}, {Title}, {Schrijver}, {Tarbell}, \& {Tomczyk}}]{2012SoPh..275..207S}
{Scherrer}, P.~H., {Schou}, J., {Bush}, R.~I., {et~al.} 2012, \solphys, 275, 207, \dodoi{10.1007/s11207-011-9834-2}

\bibitem[{{Scolini} {et~al.}(2020){Scolini}, {Chan{\'e}}, {Temmer}, {Kilpua}, {Dissauer}, {Veronig}, {Palmerio}, {Pomoell}, {Dumbovi{\'c}}, {Guo}, {Rodriguez}, \& {Poedts}}]{2020ApJS..247...21S}
{Scolini}, C., {Chan{\'e}}, E., {Temmer}, M., {et~al.} 2020, \apjs, 247, 21, \dodoi{10.3847/1538-4365/ab6216}

\bibitem[{{Sun}(2013)}]{2013arXiv1309.2392S}
{Sun}, X. 2013, arXiv e-prints, arXiv:1309.2392, \dodoi{10.48550/arXiv.1309.2392}

\bibitem[{{Sun} {et~al.}(2017){Sun}, {Hoeksema}, {Liu}, {Kazachenko}, \& {Chen}}]{2017ApJ...839...67S}
{Sun}, X., {Hoeksema}, J.~T., {Liu}, Y., {Kazachenko}, M., \& {Chen}, R. 2017, \apj, 839, 67, \dodoi{10.3847/1538-4357/aa69c1}

\bibitem[{{Sun} {et~al.}(2012){Sun}, {Hoeksema}, {Liu}, {Wiegelmann}, {Hayashi}, {Chen}, \& {Thalmann}}]{2012ApJ...748...77S}
{Sun}, X., {Hoeksema}, J.~T., {Liu}, Y., {et~al.} 2012, \apj, 748, 77, \dodoi{10.1088/0004-637X/748/2/77}

\bibitem[{{Sun} {et~al.}(2015){Sun}, {Bobra}, {Hoeksema}, {Liu}, {Li}, {Shen}, {Couvidat}, {Norton}, \& {Fisher}}]{2015ApJ...804L..28S}
{Sun}, X., {Bobra}, M.~G., {Hoeksema}, J.~T., {et~al.} 2015, \apjl, 804, L28, \dodoi{10.1088/2041-8205/804/2/L28}

\bibitem[{{Sun} {et~al.}(2023){Sun}, {Li}, {Tian}, {Hou}, {Hou}, {Chen}, {Bai}, \& {Deng}}]{2023ApJ...953..148S}
{Sun}, Z., {Li}, T., {Tian}, H., {et~al.} 2023, \apj, 953, 148, \dodoi{10.3847/1538-4357/ace5b1}

\bibitem[{{Sun} {et~al.}(2024{\natexlab{a}}){Sun}, {Li}, {Wang}, {Yang}, {Zhang}, \& {Chen}}]{2024A&A...686A.148S}
{Sun}, Z., {Li}, T., {Wang}, Q., {et~al.} 2024{\natexlab{a}}, \aap, 686, A148, \dodoi{10.1051/0004-6361/202348734}

\bibitem[{{Sun} {et~al.}(2024{\natexlab{b}}){Sun}, {Li}, {Hou}, {Tian}, {Wu}, {Li}, {Zhang}, {Li}, {Bai}, {Feng}, {Li}, {Hou}, {Song}, {Wang}, \& {Zhou}}]{2024SoPh..299...93S}
{Sun}, Z., {Li}, T., {Hou}, Y., {et~al.} 2024{\natexlab{b}}, \solphys, 299, 93, \dodoi{10.1007/s11207-024-02329-4}

\bibitem[{{Thalmann} {et~al.}(2025){Thalmann}, {Gupta}, {Veronig}, \& {Liu}}]{2025arXiv250105116T}
{Thalmann}, J.~K., {Gupta}, M., {Veronig}, A.~M., \& {Liu}, Y. 2025, arXiv e-prints, arXiv:2501.05116, \dodoi{10.48550/arXiv.2501.05116}

\bibitem[{{Vr{\v{s}}nak}(2008)}]{2008AnGeo..26.3089V}
{Vr{\v{s}}nak}, B. 2008, Annales Geophysicae, 26, 3089, \dodoi{10.5194/angeo-26-3089-2008}

\bibitem[{{Wang} \& {Liu}(2010)}]{2010ApJ...716L.195W}
{Wang}, H., \& {Liu}, C. 2010, \apjl, 716, L195, \dodoi{10.1088/2041-8205/716/2/L195}

\bibitem[{{Wang} \& {Liu}(2015)}]{2015RAA....15..145W}
---. 2015, Research in Astronomy and Astrophysics, 15, 145, \dodoi{10.1088/1674-4527/15/2/001}

\bibitem[{{Wang} {et~al.}(2016{\natexlab{a}}){Wang}, {Yan}, {Qu}, {Xue}, {Xiang}, \& {Li}}]{2016ApJ...817..156W}
{Wang}, J., {Yan}, X., {Qu}, Z., {et~al.} 2016{\natexlab{a}}, \apj, 817, 156, \dodoi{10.3847/0004-637X/817/2/156}

\bibitem[{{Wang} {et~al.}(2016{\natexlab{b}}){Wang}, {Liu}, {Wiegelmann}, {Cheng}, {Hu}, \& {Yang}}]{2016SoPh..291.1159W}
{Wang}, R., {Liu}, Y.~D., {Wiegelmann}, T., {et~al.} 2016{\natexlab{b}}, \solphys, 291, 1159, \dodoi{10.1007/s11207-016-0881-6}

\bibitem[{{Wang} {et~al.}(2012){Wang}, {Liu}, {Liu}, {Deng}, {Liu}, \& {Wang}}]{2012ApJ...745L..17W}
{Wang}, S., {Liu}, C., {Liu}, R., {et~al.} 2012, \apjl, 745, L17, \dodoi{10.1088/2041-8205/745/2/L17}

\bibitem[{{Welsch}(2022)}]{2022arXiv221101911W}
{Welsch}, B.~T. 2022, arXiv e-prints, arXiv:2211.01911, \dodoi{10.48550/arXiv.2211.01911}

\bibitem[{{Wheatland} {et~al.}(2018){Wheatland}, {Melrose}, \& {Mastrano}}]{2018ApJ...864..159W}
{Wheatland}, M.~S., {Melrose}, D.~B., \& {Mastrano}, A. 2018, \apj, 864, 159, \dodoi{10.3847/1538-4357/aad8ae}

\bibitem[{{Wyper} {et~al.}(2017){Wyper}, {Antiochos}, \& {DeVore}}]{2017Natur.544..452W}
{Wyper}, P.~F., {Antiochos}, S.~K., \& {DeVore}, C.~R. 2017, \nat, 544, 452, \dodoi{10.1038/nature22050}

\bibitem[{{Yadav} \& {Kazachenko}(2023)}]{2023ApJ...944..215Y}
{Yadav}, R., \& {Kazachenko}, M.~D. 2023, \apj, 944, 215, \dodoi{10.3847/1538-4357/acaa9d}

\bibitem[{{Yang} {et~al.}(2024){Yang}, {Tian}, {Tomczyk}, {Liu}, {Gibson}, {Morton}, \& {Downs}}]{2024Sci...386...76Y}
{Yang}, Z., {Tian}, H., {Tomczyk}, S., {et~al.} 2024, Science, 386, 76, \dodoi{10.1126/science.ado2993}

\bibitem[{{Yang} {et~al.}(2020){Yang}, {Bethge}, {Tian}, {Tomczyk}, {Morton}, {Del Zanna}, {McIntosh}, {Karak}, {Gibson}, {Samanta}, {He}, {Chen}, \& {Wang}}]{2020Sci...369..694Y}
{Yang}, Z., {Bethge}, C., {Tian}, H., {et~al.} 2020, Science, 369, 694, \dodoi{10.1126/science.abb4462}

\bibitem[{{Zhao} {et~al.}(2016){Zhao}, {Gilchrist}, {Aulanier}, {Schmieder}, {Pariat}, \& {Li}}]{2016ApJ...823...62Z}
{Zhao}, J., {Gilchrist}, S.~A., {Aulanier}, G., {et~al.} 2016, \apj, 823, 62, \dodoi{10.3847/0004-637X/823/1/62}

\bibitem[{Zuccarello {et~al.}(2017)Zuccarello, Aulanier, Dud{\'\i}k, D{\'e}moulin, Schmieder, \& Gilchrist}]{zuccarello2017vortex}
Zuccarello, F.~P., Aulanier, G., Dud{\'\i}k, J., {et~al.} 2017, The Astrophysical Journal, 837, 115

\end{thebibliography}

\begin{figure}
    \centering
    \plotone{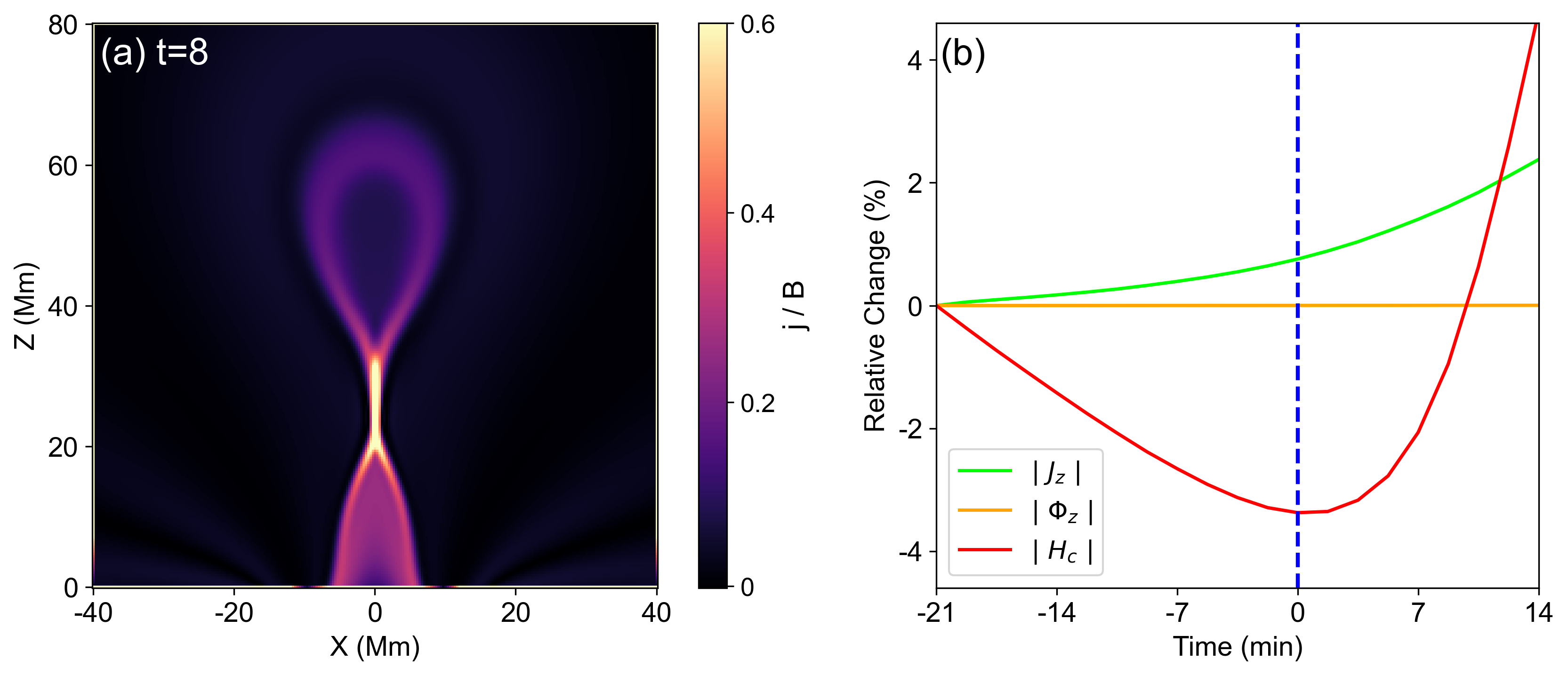}
  \caption{Overview of the MHD model (a) and the photospheric parameter evolution (b). Panel (a) shows the X-Z slice of $\mathbf{j/B}$ at t=8 (after the eruption onset). Parameters in panel (b) are calculated from all the grids in the Z=0 plane. The green, orange, and red curves indicate $J_z$, $\Phi_z$, and $H_c$, respectively. The dashed blue line indicates the time of the eruption onset ($t=0$). $t\leq 0$ indicates the pre-eruption periods.
    \label{fig:Fig.1}}
  \end{figure}

    \begin{figure}
    \centering
    \plotone{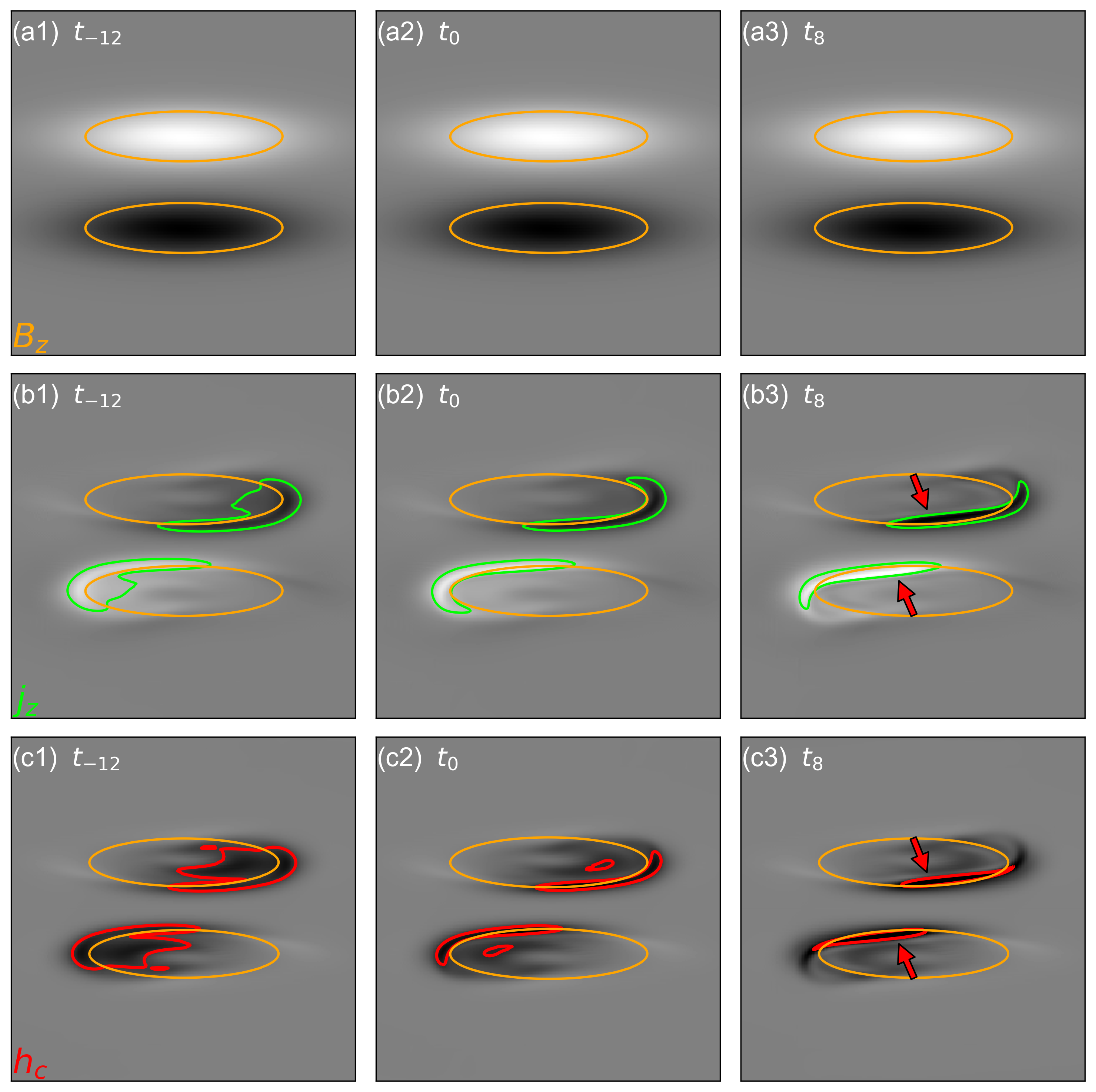}
  \caption{Evolution of the photospheric $B_z$ (a1-a3), $j_z$ (b1-b3), and $h_c$ (c1-c3) in the model. The first, second, and third column indicate the pre-eruption, eruption onset, and post-eruption states of the parameters, respectively. The orange, green, and red contour indicates the $\pm$50\% of their absolute maximum values. The red arrows mark the location where the double-J structure moves toward the center of the magnetic poles (An animation S1 of this figure is available. The animation shows the time evolution of the $B_z$,  $j_z$, and $h_c$, and base difference of $j_z$.)
    \label{fig:Fig.2}}
  \end{figure}

  \begin{figure}
    \centering
    \plotone{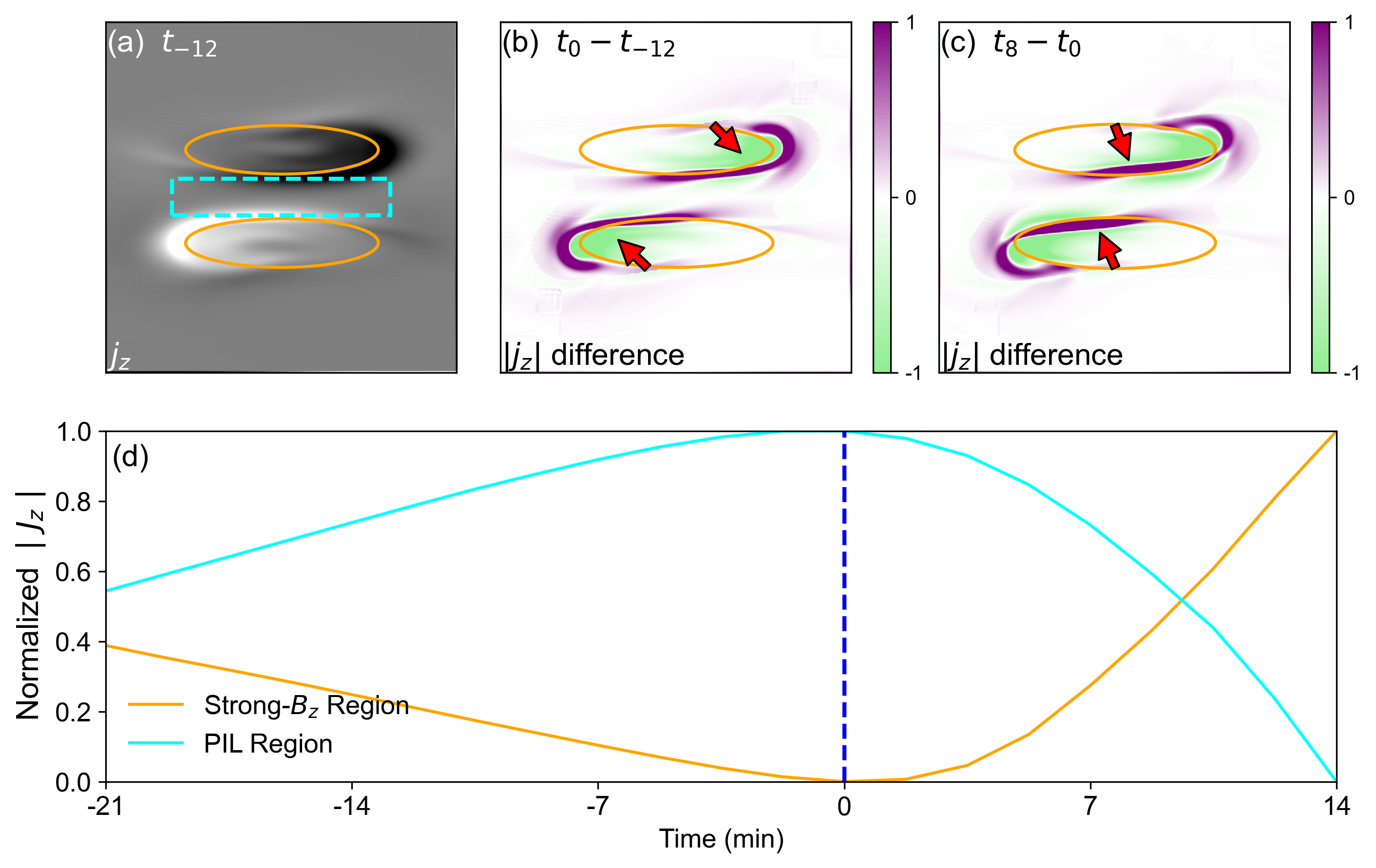}
  \caption{Evolution of the electric current density in the model. Panel (a) shows the pre-eruption distribution of $j_z$. The orange contour and the cyan dashed square mark the regions of magnetic poles and PIL, respectively. Panel (b) exhibits the $j_z$ difference between eruption onset and pre-eruption, which can indicate the pre-eruption variations of $j_z$. Panel (c) presents the $j_z$ difference between post-eruption and eruption onset, which reflects the post-eruption variations of $j_z$. The red arrows mark the region where current density variation is significant in the magnetic poles.
Panel (d) shows the average $j_z$ variations of the magnetic poles (orange) and PIL (cyan). 
    \label{fig:Fig.3}}
  \end{figure}

    \begin{figure}
    \centering
    \plotone{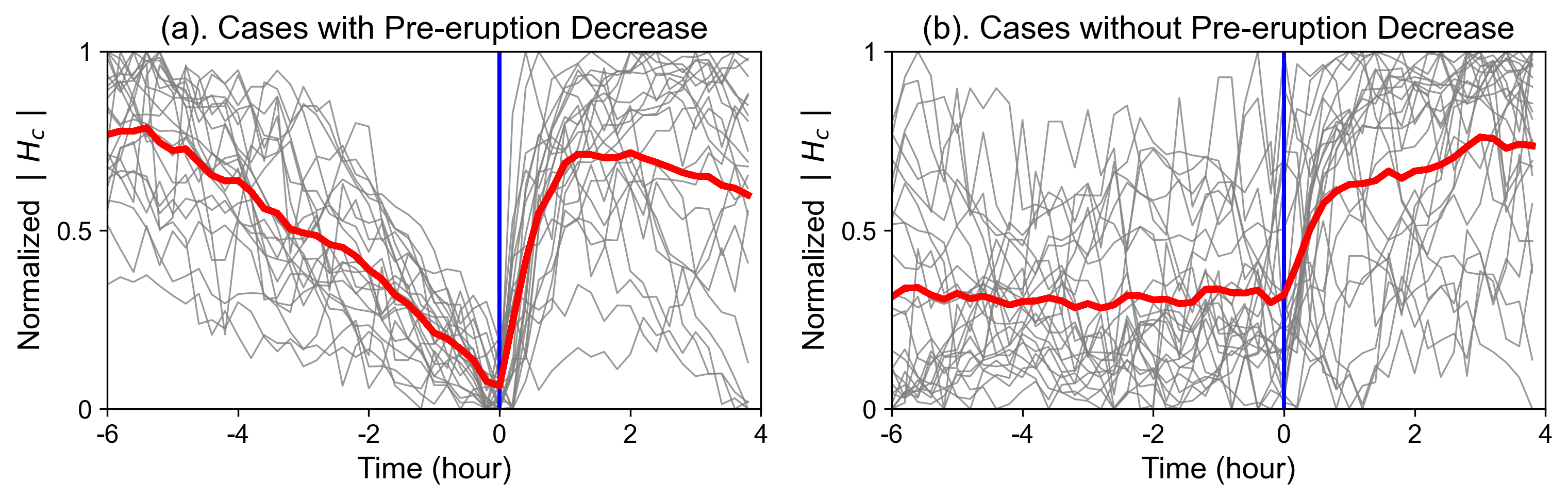}
  \caption{Current helicity evolution before and after CMEs in the observations. The samples can be divided into cases with pre-eruption decrease (a) and cases without pre-eruption decrease (b). The blue line indicates the time of the starting time of the eruption and the red lines exhibit temporal mean values of the samples.
    \label{fig:Fig.4}}
  \end{figure}

    \begin{figure}
    \centering
    \plotone{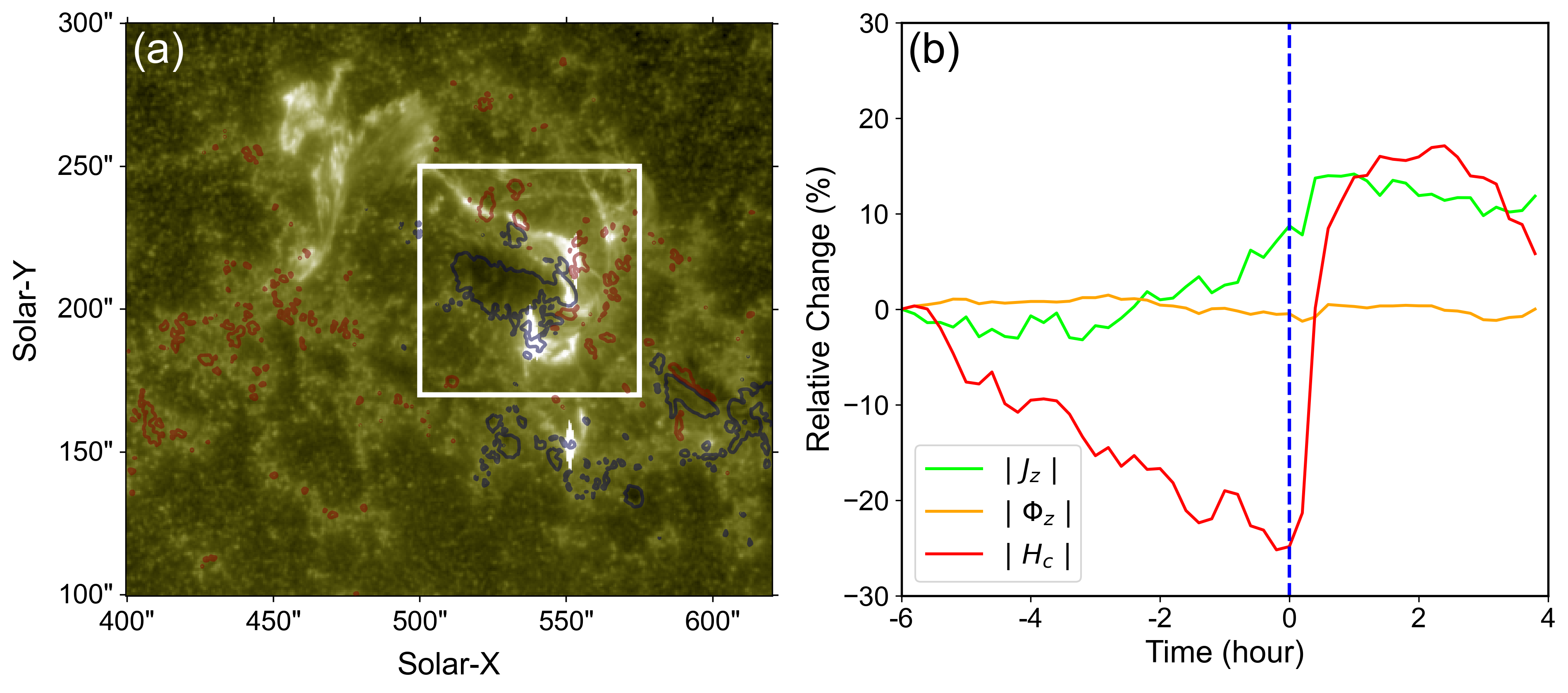}
  \caption{Overview of the eruption in AR 11261 in 1600~\AA\ (a) and the photospheric parameter evolution (b). The white square in panel (a) indicates the field of view in Figure 6. The red and blue contours indicate the positive and negative $B_z$ of 400 Gauss, respectively. The green, orange, red curves indicate $J_z$, $\Phi_z$, and $H_c$, respectively.
    \label{fig:Fig.5}}
  \end{figure}

    \begin{figure}
    \centering
    \plotone{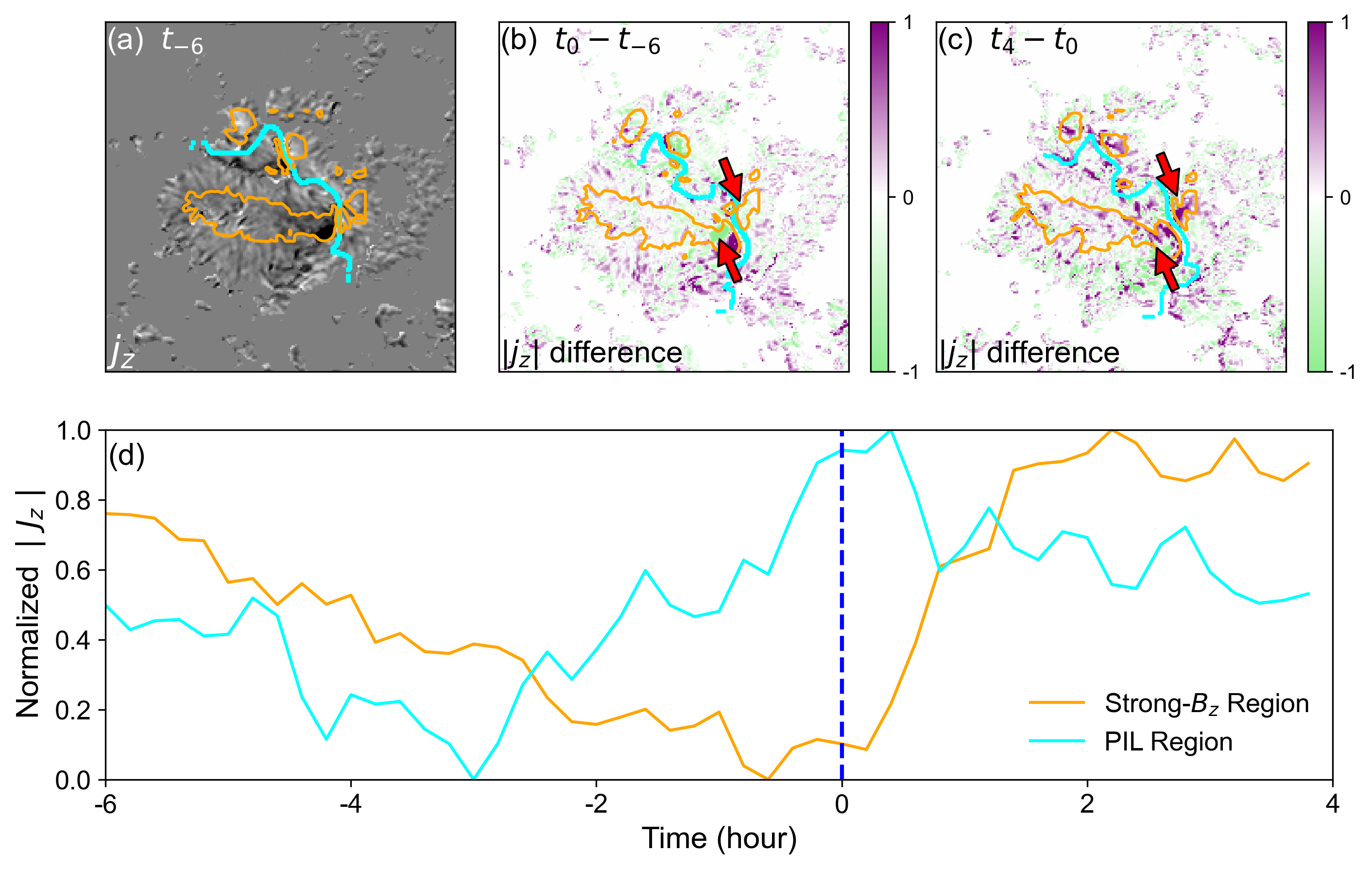}
  \caption{Evolution of the electric current density in AR 11261. Panel (a) shows the pre-eruption distribution of $j_z$. The orange contour and the blue dashed square mark the regions with 1000-Gauss $B_z$ and PIL region, respectively. The methods for calculating PIL region is referred to Section 3.1. Panel (b) exhibits the $j_z$ difference between eruption onset and pre-eruption, which can indicate the pre-eruption variations of $j_z$. Panel (c) presents the $j_z$ difference between post-eruption and eruption onset, which reflects the post-eruption variations of $j_z$. The red arrows mark the $B_z$-strong regions where the current density variation is significant.
Panel (d) shows the average $j_z$ variations of the regions with 1000-Gauss $B_z$ (orange) and PIL (cyan). (An animation S2 of this figure is available. The animation shows the time evolution of the $B_z$,  $j_z$, and $h_c$, and base difference of $j_z$.)
    \label{fig:Fig.7}}
  \end{figure}

    \begin{figure}
    \centering
    \plotone{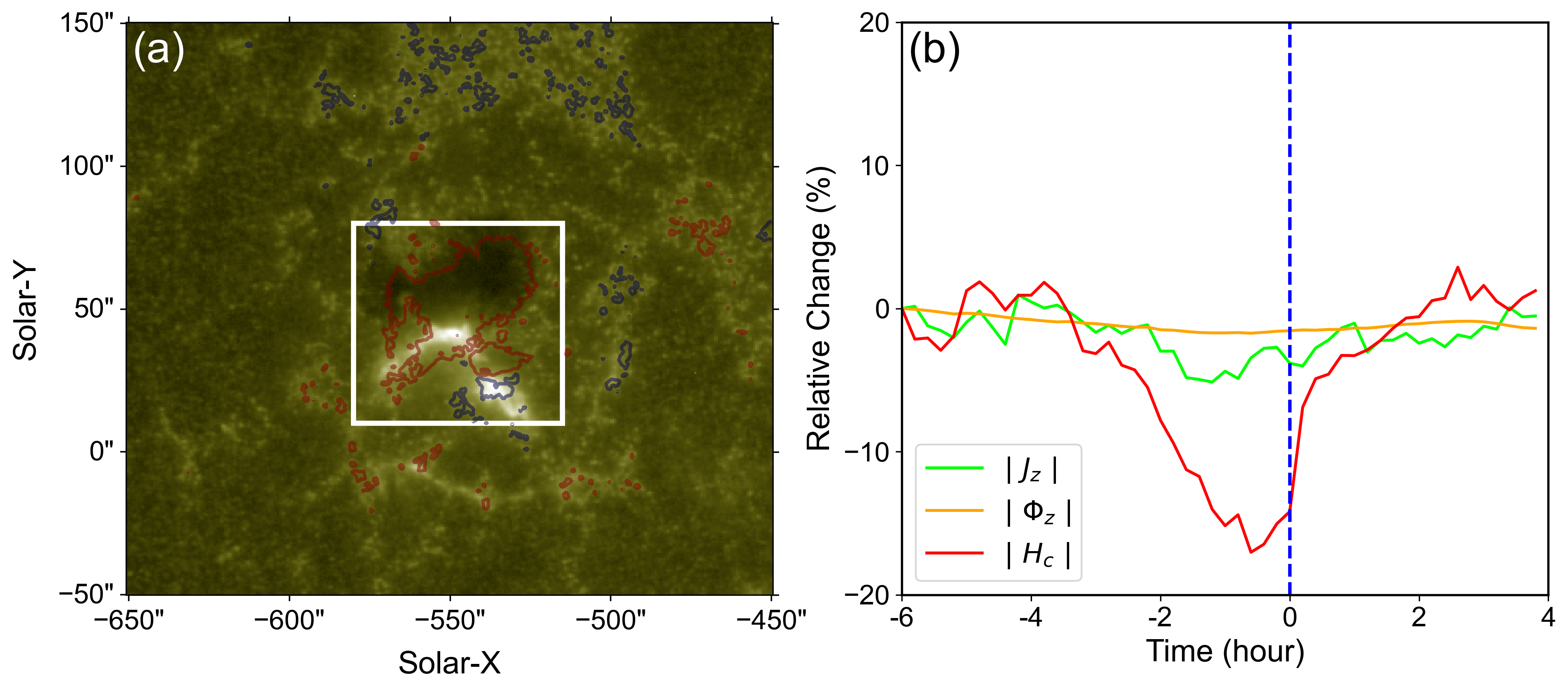}
  \caption{The same as Figure 5, but for AR 13435.
    \label{fig:Fig.8}}
  \end{figure}

    \begin{figure}
    \centering
    \plotone{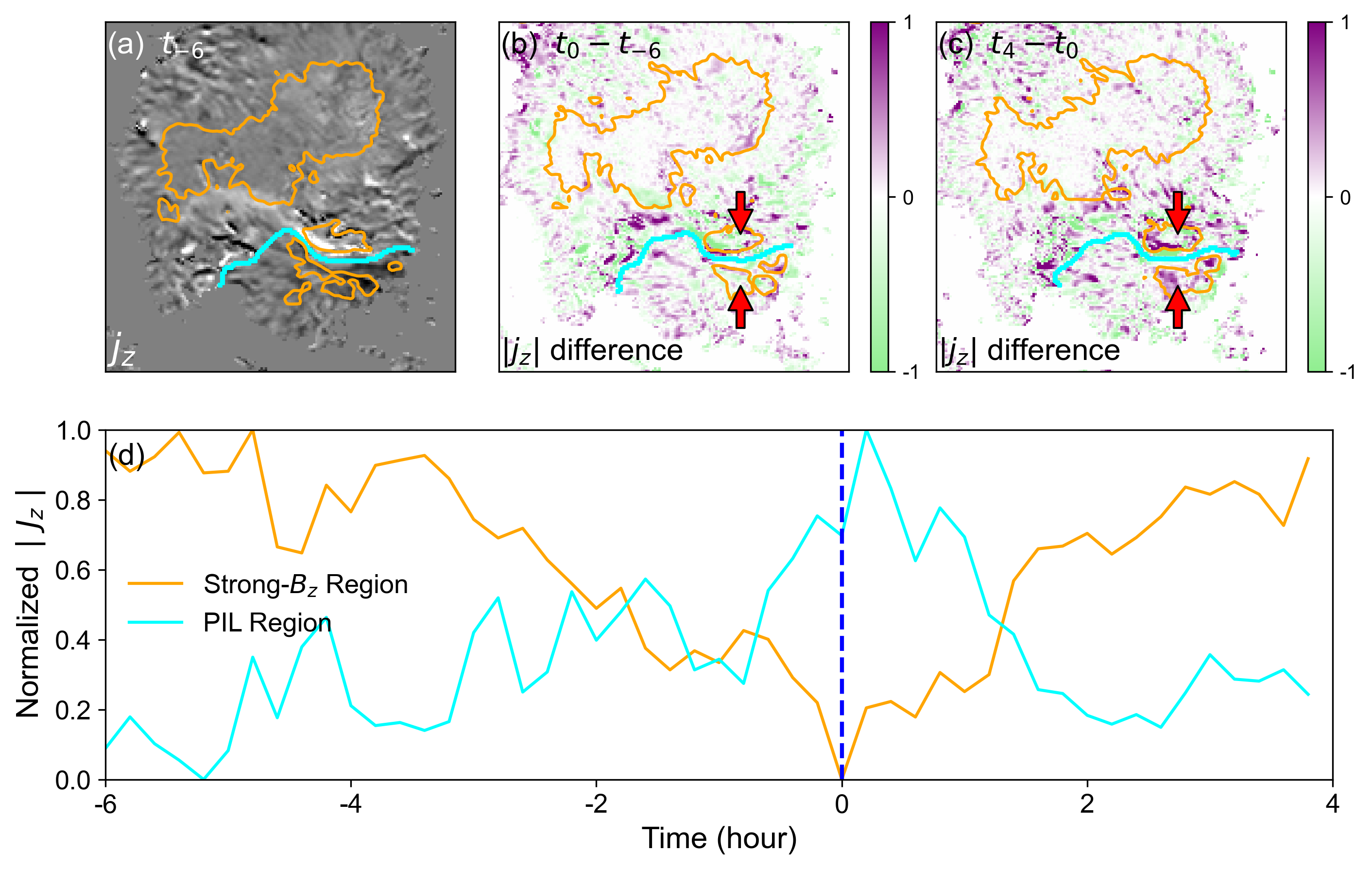}
  \caption{The same as Figure 6, but for AR 13435 (An animation S3 of this figure is available. The animation shows the time evolution of the $B_z$,  $j_z$, and $h_c$, and base difference of $j_z$.).
    \label{fig:Fig.10}}
  \end{figure}

      \begin{figure}
    \centering
    \plotone{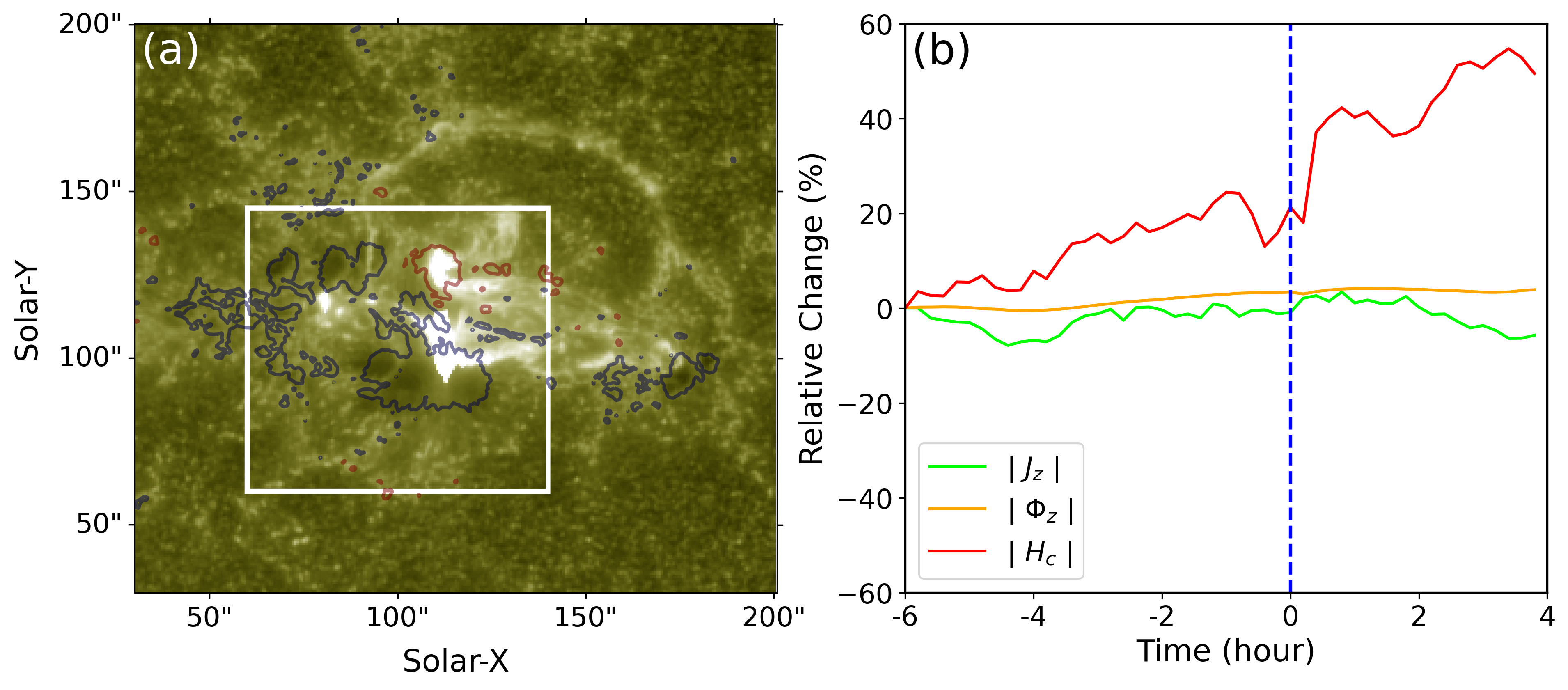}
  \caption{The same as Figure 5, but for AR 11283.
    \label{fig:Fig.9_rev}}
  \end{figure}

    \begin{figure}
    \centering
    \plotone{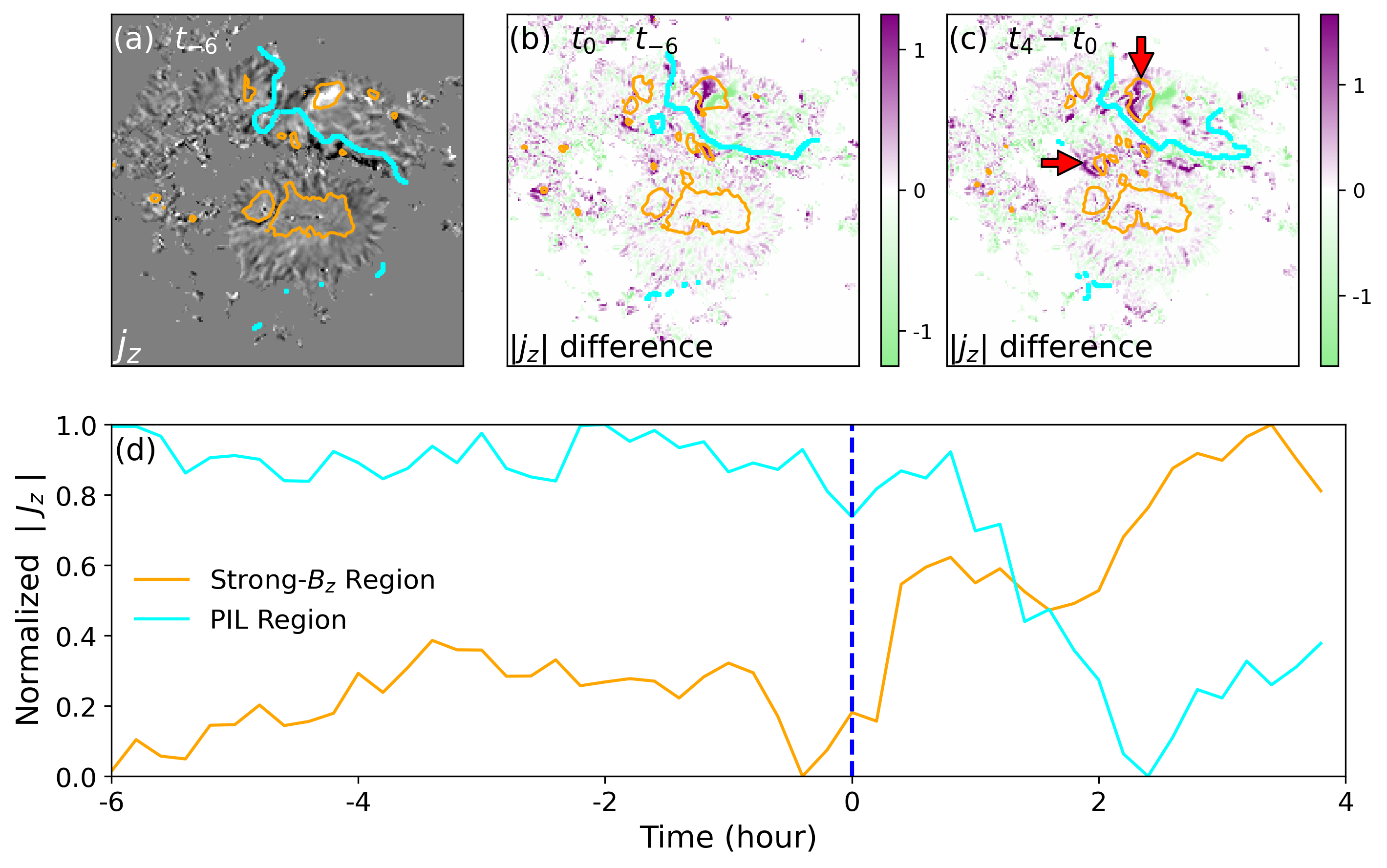}
  \caption{The same as Figure 6, but for AR 11283 (An animation S4 of this figure is available. The animation shows the time evolution of the $B_z$,  $j_z$, and $h_c$, and base difference of $j_z$.)
    \label{fig:Fig.10_rev}}
  \end{figure}

    \begin{figure}
    \centering
    \plotone{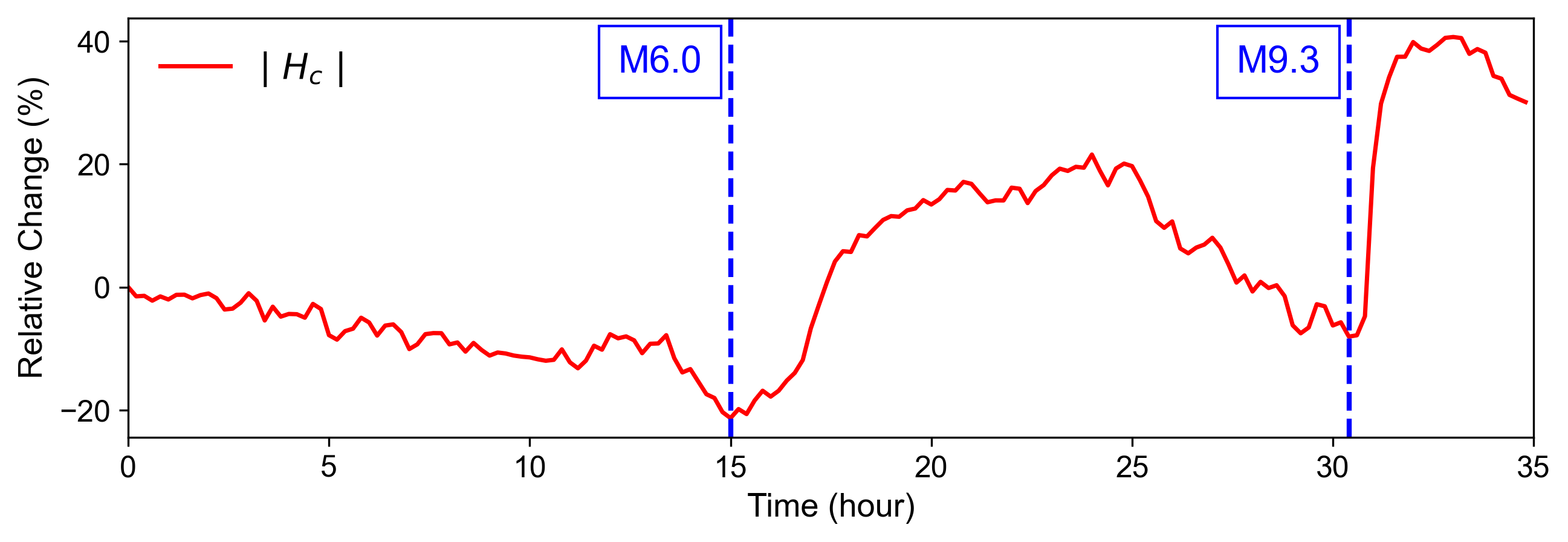}
  \caption{Long-term evolution of current helicity in AR 11261. The beginning time is at 20:36 UT on 2011 Aug 2.
    \label{fig:Fig.11}}
  \end{figure}

    \begin{figure}
    \centering
    \plotone{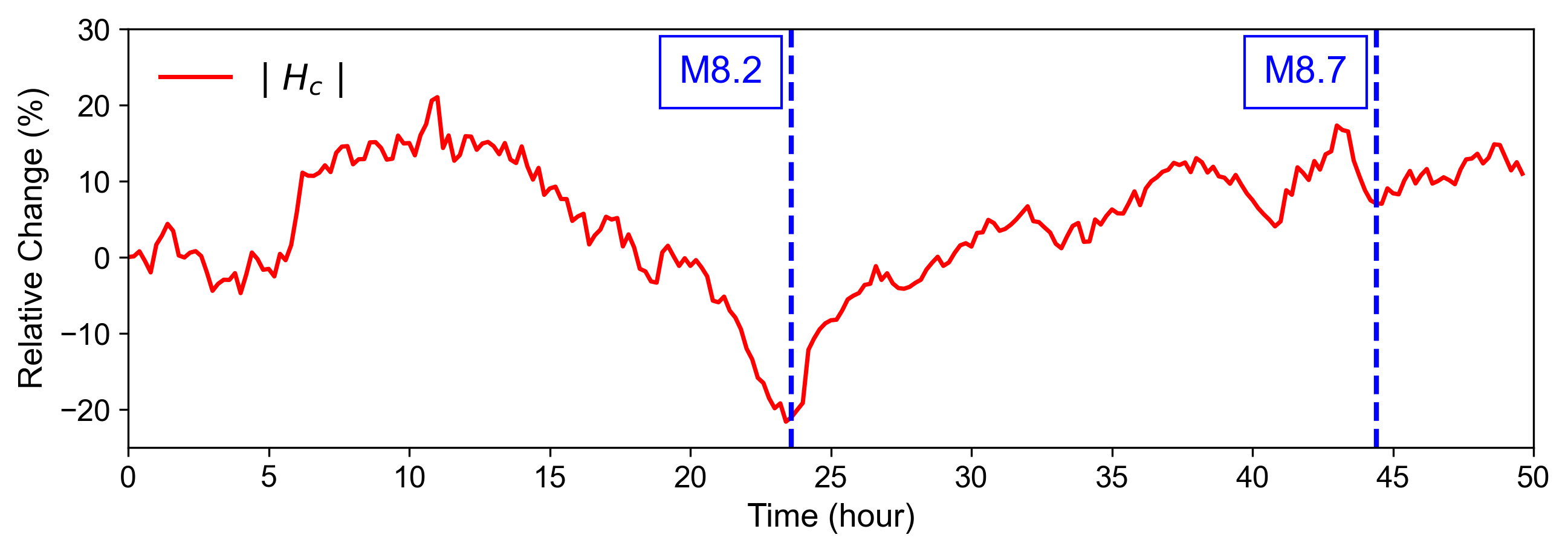}
  \caption{The same as Figure 11, but for AR 13435. The beginning time is at 14:12 UT on 2023 Sep 19.
    \label{fig:Fig.12}}
  \end{figure}

\end{document}